\documentclass[aps,prd,reprint,nofootinbib,longbibliography,preprintnumbers]{revtex4-1}
\usepackage{blindtext}
\usepackage{mathtools}
\usepackage[dvipsnames]{xcolor}
\usepackage{adjustbox}
\usepackage{siunitx}
\usepackage{appendix}
\usepackage{amssymb,amsfonts,slashed,amsthm,amsmath,graphicx,soul}

\usepackage[utf8]{inputenc}
\usepackage[english]{babel}
\usepackage{hyperref}
\hypersetup{
    colorlinks=true,
    linkcolor=blue,
    citecolor=blue,
    filecolor=blue,      
    urlcolor=blue,
}

\usepackage[normalem]{ulem}

\begin{document}

\preprint{{\rm RESCEU-17/21}}

\title{Microlensing constraints on axion stars including finite lens and source size effects}

\author{Kohei Fujikura${^1}$, Mark P. Hertzberg${^2}$,\\ Enrico D. Schiappacasse$^{3,4}$, Masahide Yamaguchi${^5}$}

\affiliation{
$^{1}$Research Center for the Early Universe (RESCEU), Graduate School of Science,
The University of Tokyo, Hongo 7-3-1 Bunkyo-ku, Tokyo 113-0033, Japan\\
$^{2}$Institute of Cosmology, Department of Physics and Astronomy, Tufts University, Medford, Massachusetts 02155, USA\\
$^{3}$ Department of Physics, University of Jyv\"{a}skyl\"{a}, P.O Box (YFL), Jyv\"{a}skyl\"{a} FI-40014, Finland\\
$^{4}$ Helsinki Institute of Physics, University of Helsinki, P.O. Box 64, Helsinki FIN-00014, Finland\\
$^{5}$Department of Physics, Tokyo Institute of Technology, Tokyo 152-8551, Japan}


\begin{abstract}
A fraction of light scalar dark matter, especially axions, may organize into Bose-Einstein condensates, gravitationally bound clumps, ``boson stars", and be present in large number in galactic halos today. We compute the expected number of gravitational microlensing events of clumps composed of the ordinary QCD axion and axionlike particles and derive microlensing constraints from the EROS-2 survey and the Subaru Hyper Suprime-Cam observation. We perform a detailed lensing calculation, including the finite lens and source size effects in our analysis. We constrain the axion mass in terms of the fraction of dark matter collapsed into clumps, the individual clump densities, and the axion self-coupling. We also consider and constrain clumps composed of a generic scalar dark matter candidate with repulsive self-interactions. Our analysis opens up a new window for the potential discovery of dark matter.
\end{abstract}

\maketitle

\renewcommand{\thefootnote}{\arabic{footnote}}
\setcounter{footnote}{0}



\section{Introduction}

Several astrophysical observations, such as galactic rotation curves, cosmic microwave background and large scale structure, are well explained by cold dark matter~\cite{Peebles:2013hla}.
Although there are a lot of several well-motivated dark matter candidates, the particle physics origin of cold dark matter is currently unknown.
Among them, the axion motivated by the solution to the strong CP problem~\cite{PhysRevLett.38.1440, PhysRevLett.40.223, PhysRevLett.40.279} of quantum chromodynamics (QCD), and axionlike particles whose existence are predicted in string theory~\cite{Svrcek:2006yi}, are prominent cold dark matter candidates.
Only a small part of the most highly motivated region of the axion’s parameter space has been probed
experimentally, but several interesting experiments have been proposed and/or planned for the incoming years (see, for example, Refs.~\cite{Irastorza:2018dyq, MADMAX:2019pub,
    Arza:2019nta, Nurmi:2021xds, Edwards:2020afl} for new experimental approaches or ways to test an axion, including indirect searches).

Axions are produced at high occupancy in the early Universe by a misalignment mechanism~\cite{Preskill:1982cy,Abbott:1982af,Dine:1982ah}.
Since a total number of produced axions is approximately conserved due to the extremely small coupling, axions form the Bose-Einstein condensate (BEC) if they are in thermal equilibrium.
Thermalization of axions would be driven by the gravitational interaction and the possibility of axion BEC is investigated in many literatures~\cite{Erken:2011vv,Saikawa:2012uk,Davidson:2013aba,Noumi:2013zga,Guth:2014hsa}.
Different from a conventional BEC, axion BEC has a short range order driven by the attractive gravitational interaction as pointed out by authors of Ref.~\cite{Guth:2014hsa}.
As a result, axion BEC forms gravitationally bound objects called \textit{axion clumps}~\cite{Guth:2014hsa,Schiappacasse:2017ham} whose configuration can be adequately captured by classical field theory as shown in Ref.~\cite{Hertzberg:2016tal}.
In the literature, these clumps are sometimes called axion stars or boson stars
(we use the words \textit{stars} and \textit{clumps} interchangeably in this paper).
Previous work includes Refs.~\cite{Tkachev:1986tr,Gleiser:1988rq,Seidel:1990jh,Tkachev:1991ka,Jetzer:1991jr,Liddle:1992fmk,Kolb:1993zz,Sharma:2008sc,Chavanis:2011zi,Chavanis:2011zm,Liebling:2012fv,Visinelli:2017ooc,Hertzberg:2018lmt}. (There has also been related work on complex scalars, including Refs.~\cite{Colpi:1986ye, Schunck:2003kk, Choi:2019mva, Guerra:2019srj, Hertzberg:2020xdn}.)
Axion clumps would typically form in the scenario where the PQ symmetry breaking takes place after the inflation (postinflationary scenario).
In this case, the axion field remains inhomogeneous from one Hubble patch to the next by causality after the PQ symmetry breaking. In such conditions, when the axion field becomes massive during the QCD phase transition, the already present axion fluctuations would begin to interact among them via strong gravitational mode-mode interactions, and eventually, axion clumps are formed after thermalization~\cite{Guth:2014hsa}.

In the scenario where the PQ symmetry breaking takes place before or during inflation (preinflationary scenario), the axion field is driven to be highly homogeneous on large scales, and thus, it is unclear if the axion may form a BEC in the late Universe.
However, some of us of the present paper pointed out in Ref.~\cite{Hertzberg:2020hsz} that the nucleation of clumps composed of QCD axion or axionlike particles may occur in dark matter minihalos around primordial black holes (PBHs).
Axion minihalos would satisfy the necessary conditions for kinetic formation of axion clumps via gravitational condensation in the so-called kinetic regime. In this regime, the length scale of the system is much longer than the wavelength of the axion field. The relaxation rate is given by \cite{Levkov:2018kau} $\Gamma_{\text{kin}} \sim n_{\phi}\,\sigma_{\text{gr}}v_{\phi}\,\mathcal{N}$, where $\sigma_{\text{gr}}$ is the gravitational scattering cross section [while the contribution from self-interactions arises from the replacement $\sigma_{\text{gr}}\to\sigma_{\text{si}}$ (the cross section of the self-interactions), which is normally negligible], $\mathcal{N}$ is the occupancy number associated with the Bose enhancement, $n_{\phi}$ is the axion number density, and $v_{\phi}$ is the typical speed of axions in minihalos.

In both scenarios, there is a constraint on the PQ symmetry breaking scale $F_a$.
In the postinflationary scenario, the decay of topological defects critically affects the axion abundance leading to the so-called domain wall problem. To avoid such a problem, it is natural to consider a domain wall number equal to the unity so that the QCD axion may explain the dark matter of the Universe in the mass range  $10^{-4}\,\text{eV}\lesssim m_{a} \lesssim 10^{-2}\,\text{eV}$, e.g., a range for the axion decay constant of $10^{9}\,\text{GeV}\lesssim  F_a \lesssim 10^{11}\,\text{GeV}$~\cite{Kawasaki:2014sqa}.
In the preinflationary scenario, where the PQ symmetry is broken before or during inflation, the domain wall problem is automatically solved by the exponential expansion of the Universe, and the axion abundance is dominated by the misalignment mechanism.
If the initial misalignment angle is the order of unity, the axion decay constant is bounded from above as $F_a \lesssim 10^{12}\,\text{GeV}$ to avoid the overclosure of the Universe.
Combining with the lower bound on $F_a$ from the observation of neutrino burst duration of SN1987A~\cite{Mayle:1987as,Raffelt:1987yt,Turner:1987by}, the constraint is given by $10^8\,{\rm GeV}\lesssim F_a \lesssim 10^{12}\,{\rm GeV}$ called the QCD axion window.
However, if an additional fine tuning is allowed in the frame of the axion anthropic window~\cite{PhysRevLett.52.1725, LINDE199138, Wilczek:2004cr, PhysRevD.73.023505}, the axion decay constant may take much larger values as suggested by unification ideas.

Axion dark matter clumps offer several ways for dark matter indirect searches, such as the collapse and explosion in relativistic axions of critical-mass axion clumps~\cite{Levkov:2016rkk} or the resonance of photons after the merger of axion clumps~\cite{Hertzberg:2018zte,Hertzberg:2020dbk}.
In addition to these searches, it has been well known that such a massive compact object can cause an amplification of brightness of a background source star when it passes close to a line of sight to that star, called gravitational lensing events.
For example, abundances of a massive astrophysical compact halo objects (MACHOs) and PBHs are stringently constrained by gravitational lensing events such as EROS/MACHO survey~\cite{MACHO:1998qtf,Tisserand:2006zx}, the Optical Gravitational Lensing Experiment (OGLE)~\cite{Udalski:1994hn,2015AcA....65....1U,Niikura:2019kqi} and the Subaru Hyper Suprime-Cam (HSC) observation~\cite{Niikura:2017zjd}.

A main purpose of the present paper is to derive microlensing constraints coming from these surveys on axion clumps.
Axion clumps have an internal structure, and hence, they generally cannot be considered as pointlike massive objects for microlensing events. This differs from the case of MACHOs and PBHs.
Thus, in order to correctly derive microlensing constraints on axion clumps, one needs to study effects on gravitational lensing from the finite extent of axion clumps (the finite lens size effect).
There are several studies of microlensing events caused by astrophysical objects which possess finite extent.
For example, gravitational lensing constraints on extended compact objects such as boson stars and self-similar subhalos are investigated in Refs~\cite{Croon:2020wpr,Croon:2020ouk}. With respect to axion dark matter substructures, gravitational lensing of axion miniclusters is studied in Refs.~\cite{Kolb:1995bu, Fairbairn:2017sil, Fairbairn:2017dmf}.
The authors of Ref.~\cite{Marfatia:2021twj} investigate microlensing constraints on fermi-balls.

In particular, we find that when the size of axion clumps is longer than the typical length scale of microlensing, which is Einstein ring radius, axion clumps cannot be considered as a pointlike massive object.
(See Sec.~\ref{sols} and Sec~\ref{sec:point lens point source} for definitions of a size of axion clumps and the Einstein ring radius, respectively.)
Resultant magnifications of source stars are significantly suppressed due to the extent of axion clumps, and hence, microlensing constraints become weak even if masses of axion clumps are sufficiently heavy so that microlensing events are triggered.
We perform numerical calculations of an expected number of microlensing events in the EROS-2 survey and Subaru HSC observation including finite lens and finite source size effects.
Microlensing constraints on clumps composed of the ordinary QCD axion, axionlike particles, and the generic light scalar fields with repulsive self-interactions are clarified.
It turns out that observations of microlensing events cannot constrain the traditional QCD axion window due to the significant finite source size effect, but a higher breaking scale $F_a\gtrsim 10^{12}\, {\rm GeV}$ can be constrained.
Recently, authors of Ref.~\cite{Sugiyama:2021xqg} also focus on axion clumps and clarify the allowed parameter space leading to the microlensing events reported by Subaru HSC and OGLE observations.

The outline of this paper is as follows.
 In Sec.~\ref{sec:review}, we describe the basics of axion dark matter clumps with a spherical symmetry. We discuss the parameter space of solutions in the nonrelativistic regime and current abundance of axion clumps in galactic halos.
In Sec. \ref{sec:microlensing}, we briefly review the basics of gravitational microlensing and calculate threshold impact parameters of clumps with the inclusion of finite lens and source size effects. In Sec.~\ref{sec:event rate and microlensing constraints}, we calculate the allowed region in the axion parameter space by estimating the expected number of microlensing events using the data obtained by EROS-2 survey and the Subaru HSC observation. In Sec.~\ref{repulsive}, we discuss microlensing constraints of clumps composed of a generic light scalar dark matter candidate with repulsive self-interactions.  Sec.~\ref{sec:conclusion} is devoted to the conclusion.
Finally, in the Appendix, we derive the lens equation including the finite lens size effect.

\section{Axion stars}\label{sec:review}

In this section, we review the dynamics of axion clumps which is relevant for this work. 
For a general review about axions and axionlike particles as dark matter particles, see Refs.~\cite{Duffy_2009, Masso:2002ip, Marsh:2015xka, Fortin:2021cog} for examples.

\subsection{Axion stars with a spherical symmetry}\label{sols}
In this subsection, we briefly review general features of axion dark matter clumps with a spherical symmetry. These gravitationally bounded astrophysical objects were studied in detail in Refs.~\cite{Chavanis:2011zi,Chavanis:2011zm,Schiappacasse:2017ham}.

In the effective theory for axions, the Lagrangian density of an axion field is given by~\footnote{Here, we use natural units ($\hbar=c=1$) and the metric signature (+ - - -).}
\begin{align}
&\mathcal{L} = \sqrt{-g} \left( \frac{1}{2}g^{\mu\nu} \nabla_\mu \phi \nabla_\nu \phi - V(\phi) \right), \label{eq:Lagrangian}
\end{align}
where $V(\phi)$ is the scalar potential of a real scalar field $\phi$ representing an axion.
It was shown in Ref.~\cite{diCortona:2015ldu} that $V(\phi)$ can be calculated by integrating out the neutral pion.
The resultant axion potential takes the following form:
\begin{align}
V(\phi)= \Lambda^4 \left[ 1-\sqrt{1-\frac{4m_u m_d}{(m_u + m_d)^2} \sin^2\left(\dfrac{\phi}{2F_a}\right)}\right]\,\textcolor{blue}{,} \label{eq:axion precisely}
\end{align}
where $m_u\simeq 2.2\,{\rm MeV},~m_d\simeq 4.7\,{\rm MeV}$, and $F_a$ are the up and down quark masses and the axion decay constant, respectively.
The overall scale of the potential, $\Lambda$, is given by~\cite{diCortona:2015ldu}
\begin{align}
    \Lambda^4 = f_\pi^2 m_\pi^2,
\end{align}
where $f_\pi\simeq 92\,{\rm MeV}$ and $m_\pi\simeq 135\,{\rm MeV}$ are the pion decay constant and the neutral pion mass, respectively.
On the other hand, for axionlike particles, we treat $\Lambda$ as a free parameter in the following discussion.

At a small field values, e.g., $\phi/F_a \ll 1$, we can expand the sine function in Eq.~\eqref{eq:axion precisely} to obtain~\footnote{One should note that the axion potential is usually calculated by assuming dilute gas approximation, which leads to $V(\phi)=\Lambda^4\left[1 -\text{cos}\left( \phi/F_a \right) \right]$.
By expanding the cosine function in this expression, the resulting axion potential takes the same form as Eq.~\eqref{eq:approximate axion potential}, with $m_a^2\equiv \Lambda^4/F_a^2$ and $\lambda\equiv m_a^2/F_a^2$.
}
\begin{align}
V (\phi ) =\frac{1}{2}m_a^2 \phi^2 - \frac{\lambda}{4!}\phi^4+\mathcal{O}\left(\frac{\phi^6}{F_a^6}\right) \label{eq:approximate axion potential},
\end{align}
where
\begin{align}
m_a^2 \equiv \frac{m_u m_d}{(m_u+m_d)^2} \frac{f_\pi^2 m_\pi^2}{F_a^2},~\lambda \equiv \gamma \frac{m_a^2}{F_a^2}\equiv \frac{m_a^2}{F_a^{\prime2}},
\end{align}
and $\gamma \equiv 1-3m_u m_d/(m_u+m_d)^2\simeq 0.34$. 
Note that we have absorbed the parameter $\gamma$ into the axion decay constant by defining $F'_a = F_a/\gamma^{1/2}$.

Let us next derive a spherically symmetric localized configuration for the field $\phi$.
In the nonrelativistic regime, it is convenient to express the real scalar field $\phi$ in terms of a complex scalar field $\psi (\bold{x},t)$ as 
\begin{align}
\phi (\bold{x},t)= \frac{1}{ \sqrt{2m_a} } \left( e^{-im_a t} \psi (\bold{x},t) +e^{im_a t}\psi^* (\bold{x},t) \right).
\end{align}
Here, $\psi (\bold{x},t)$ is a slowly varying function satisfying $|\dot{\psi}(\bold{x},t)/m_a| \ll |\psi(\bold{x},t)|$.
Using the weak field gravitational approximation, the dynamics of $\psi$ is governed by the following non-relativistic Hamiltonian~\cite{Schiappacasse:2017ham},
\begin{align}
H_{\rm tot}\equiv H_{\rm kin} + H_{\rm int}+H_{\rm gravity}\,, \label{eq:total Hamitonian}
\end{align}
where
\begin{align}
&H_{\rm kin} = \frac{1}{2m_a}\int d^3 {\bm x} \nabla \psi^*({\bm x}) \nabla \psi({\bm x}),\nonumber\\
&H_{\rm int}=-\frac{\lambda}{16m_a^2} \int d^3 {\bm x} |\psi ({\bm x})|^4, \\
&H_{\rm gravity}=-\frac{G_N m_a^2}{2}\int d^3 {\bm x} \int d^3 {\bm x}' \frac{|\psi({\bm x})|^2 |\psi ({\bm x}')|^2}{|{\bm x}-{\bm x}'|}\nonumber.
\end{align}
Here, $G_N$ is the Newtonian constant of gravitation as usual, and  $H_{\rm kin},~H_{\rm int}$, and $H_{\rm gravity}$ are the kinetic energy, the self-interacting energy, and the gravitational energy of an axion configuration, respectively.
The above Hamiltonian possesses a global ${\rm U(1)}$ symmetry whose transformation is defined by $\psi\to e^{i\theta}\psi$.
This global symmetry implies that the total number of axions $N$ is conserved, which was defined as
\begin{align}
N\equiv \int d^3 x |\psi|^2\,. \label{eq:axion number}
\end{align}
 This conserved quantity ensures the stability of the axion clump from the viewpoint of particle-number violated processes, which are usually highly suppressed in the nonrelativistic limit.~\footnote{However, there may be present unavoidable number changing processes through the coupling between the axion and photons, as is discussed by some of us in Refs.~\cite{Hertzberg:2018zte, Hertzberg:2020dbk}. We comment on this effect later.} 

\begin{figure*}[t]
\centering\includegraphics[width=15cm]{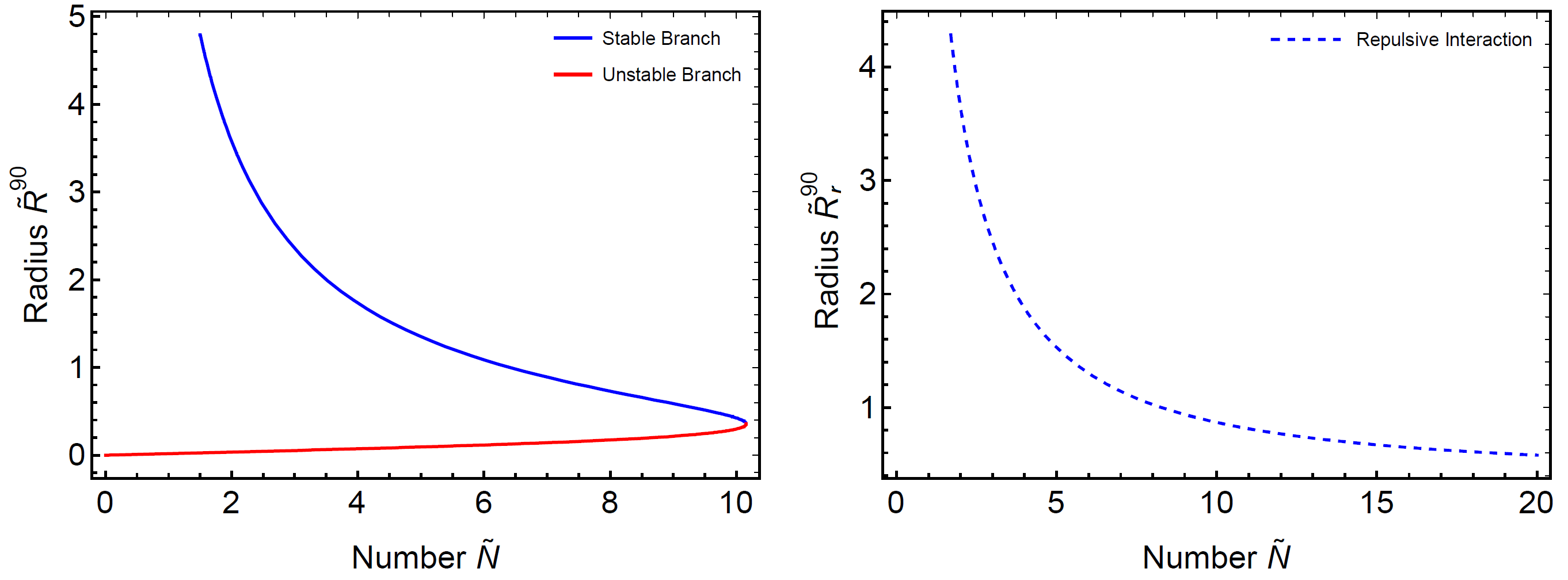}
\caption{
(Left) Two branches of solutions for the axion clump in the parameter space given by the dimensionless radius $\tilde{R}^{90}$ (which encloses the $90\%$ of the total mass) and the dimensionless number of particles $\tilde{N}$. While the upper blue curve corresponds to stable solutions under radial perturbations, the lower red curve refers to unstable solutions. Both solutions are obtained using the exponential $\times$ linear ansatz, Eq.~(\ref{eq:ansatz}), where $\tilde{R}^{90}\approx 3.610 \tilde{R}$. (Right) Single (stable) branch of solutions for the case of generic scalar dark matter with repulsive self-interactions (see Sec.~\ref{repulsive}). The subscript $r$ in $\tilde{R}_r^{90}$ just indicates the flipped sign in the expression for the length scale, Eq.~(\ref{eq:radius of the clump}), as explained in the main text.  In both panels, the system is treated in the nonrelativistic regime.
}\label{fig:branches}
\end{figure*}

In this paper, we are mainly interested in spherically symmetric axion clumps, which correspond to a true BEC defined by a fixed number of particles.~\footnote{The inclusion of nonzero angular momentum leads to higher eigenstates of the axion condensate. These configurations may contain a larger number of particles than the corresponding spherically symmetric clumps. This feature enhances the possibility that rotating clumps undergo parametric resonance of photons under a suitable value for the 
axion-photon coupling constant~\cite{Hertzberg:2018lmt, Hertzberg:2018zte}.} A spherically symmetric stationary configuration
can be obtained by expressing the complex field $\psi({\bf{x}})$ as
\begin{equation}
\psi(r,t) = \Psi(r) e^{-i\mu t}\,,
\end{equation}
where $r = |\bf{x}|$, and $\mu$ is much smaller than the axion mass
as expected in the nonrelativistic approximation, and $\Psi(r)$ is the clump radial profile.
It was shown in Ref.~\cite{Schiappacasse:2017ham} that this radial profile can be well fitted by an exponential $\times$ linear ansatz according to
\begin{align}
\Psi(r) = \sqrt{\frac{N}{7\pi R^3}}\left(1+\frac{r}{R}\right) e^{-r/R},\label{eq:ansatz}
\end{align}
where the prefactor $\sqrt{N/(7\pi R^3)}$ ensures the normalization given by Eq. \eqref{eq:axion number}, and $R$ is the length scale which controls the shape of the profile.
For later convenience, let us rewrite the clump number of particles, the length scale and the total Hamiltonian of the system in their respective dimensionless correspondences as follows:
\begin{align}
&\widetilde{N} \equiv \frac{m_a^2 \sqrt{G_N}}{F_a'} N, \label{eq:particle numbers}\\
&\widetilde{R} = m_a \sqrt{G_N} F_a'R,\\
&\widetilde{H} \equiv \frac{m_a}{F_a'^3\sqrt{G_N}}H\,.
\end{align}
With the ansatz given in Eq.~\eqref{eq:ansatz}, the total Hamiltonian in Eq.~\eqref{eq:total Hamitonian} can be rewritten as
\begin{align}
\widetilde{H}_{\rm tot}= a \frac{\widetilde{N}}{\widetilde{R}^2} -b \frac{\widetilde{N}^2}{\widetilde{R}}-c\frac{\widetilde{N}^2}{\widetilde{R}^3}, \label{eq:Hamiltonian}
\end{align}
where
\begin{align}
a=\frac{3}{14},~b=\frac{5373}{25088},~c=\frac{437}{200704\pi}. \label{eq:constants}
\end{align}
Extremizing $H_{\rm tot}$ with respect to the length scale $\widetilde{R}$ at a fixed number of particles, we obtain the parameter space of solutions for the axion self-gravitating system in the nonrelativistic approximation according to
\begin{align}
\widetilde{R}=  \frac{a\pm \sqrt{a^2-3bc\widetilde{N}^2} }{b\widetilde{N}}. \label{eq:radius of the clump}
\end{align}
There are two branches of solutions in the $(\tilde{N},\widetilde{R})$ space as shown in Fig.~\ref{fig:branches} (left panel)~\footnote{Note that for $a^2-3bc\widetilde{N} <0$, the negative energy from attractive self-interactions and the Newton potential dominate over entire $\widetilde{R}$ region giving rise to the absence of the extremum of $\widetilde{H}_{\rm tot}$.}.
Keeping the negative sign in front of the square root in Eq.~\eqref{eq:radius of the clump} leads to a branch in which the axion attractive self-interaction dominates over gravity (red curve).~\footnote{ For very small axion clumps, higher order terms of the potential $V(\phi)$ eventually become important, and the nonrelativistic approximation breaks down. A relativistic treatment in this regime shows the presence of new solutions called axitons. These clumps are short lived due to the emission of relativistic axions~\cite{Schiappacasse:2017ham, Kolb:1993hw}.} Clumps on this branch are unstable under small radial perturbations. By contrast, keeping the positive sign instead of the negative one leads to a branch in which gravity dominates over the axion self-interaction and clumps are stable (blue curve). 

For completeness, we also show in Fig.~\ref{fig:branches} (right panel) the case of a generic light scalar dark matter holding a quartic repulsive self-interaction. We analyze such system in detail in Sec.~\ref{repulsive}. After suitable transformations, the dimensionless Hamiltonian of any localized (spherically symmetric) clump ansatz is given by Eq.~(\ref{eq:Hamiltonian}), but with the sign of the self-interacting term (that is, the coefficient $c$) flipped. As a result, there is only one (stable) branch of solutions as the dashed blue curve shows in Fig.~\ref{fig:branches} (right panel).

Now, we return to the case of the attractive interactions, which is the main topic in this paper since it applies to axions. The upper branch is stable under radial perturbations \cite{Schiappacasse:2017ham}. So the clumps on this branch are astrophysical objects that we focus on, as they can give rise to gravitational lensing effects thanks to their large masses.
Both branches of solutions converge into a point which corresponds to the clump with a maximum number of particles, $\widetilde{N}_{\text{max}}$, having the minimum size,
$\widetilde{R}_{\text{min}}$. In this point, gravity and the axion self-interaction are comparable. For the exponential $\times$ linear ansatz that we are using, we have
\begin{align}
\widetilde{N}_{\rm max} = \frac{a}{\sqrt{3bc}}\simeq 10.2, \label{eq:Nmax}\\
\widetilde{R}_{\rm min}= \frac{a}{b\widetilde{N}_{\rm max}}\simeq 0.098.
\end{align}
By using $\widetilde{N}_{\rm max}$ and $\widetilde{R}_{\rm min}$, we can express generic $\widetilde{N}$ and $\widetilde{R}$ in terms of one parameter $\alpha$ as
\begin{align}
\widetilde{N}=\alpha \widetilde{N}_{\rm max},\hspace{1 cm}~\widetilde{R} = \frac{1}{\alpha} \widetilde{R}_{\rm min} \left(1+\sqrt{1-\alpha^2}\right)\,, \label{eq:alpha definition}
\end{align}
where $0<\alpha<1$.
For the QCD axion and axionlike particles, the typical total number of particles, size, and mass of clumps are estimated as
\begin{align}
N 
\simeq 1.7\times10^{60} \times \alpha &\left( \frac{10^{-5} {\rm eV}}{m_a}\right)^2  \nonumber \\
&\times \left( \frac{F_a}{10^{12} {\rm GeV}} \right)\left(\frac{0.3}{\gamma}\right)^{\frac{1}{2}} ,\label{Neq}
\end{align}
\begin{align}
R\simeq1.8\times 10^4\,{\rm m}  \times &\left(\frac{1+\sqrt{1-\alpha^2}}{\alpha}\right) \left( \frac{10^{-5}{\rm eV}}{m_a}\right)\nonumber \\
 &\times\left( \frac{10^{12} {\rm GeV}}{F_a}\right)\left(\frac{\gamma}{0.3}\right)^{\frac{1}{2}}  ,\label{Req}
\end{align}
\begin{align}
M_{\rm clump}=Nm_a \simeq 
 1.5&\times 10^{-11}M_{\odot}\times \alpha \left( \frac{10^{-5} {\rm eV}}{m_a}\right) \nonumber \\
 &\times \left( \frac{F_a}{10^{12} {\rm GeV}} \right) \left(\frac{0.3}{\gamma}\right)^{\frac{1}{2}} \label{Meq}, 
\end{align}
where $M_{\odot}$ is the solar mass.
For the QCD axion, we have $m_a = 10^{-5}\,\text{eV}(6\times 10^{11}\,\text{GeV}/F_a)$ in the above equations. Note that when $\alpha = 1$, we have the maximum number of particles, the minimum length scale, and the maximum mass for a spherically symmetric axion clump in Eqs.~\eqref{Neq}, \eqref{Req}, and \eqref{Meq}, respectively.
By imposing the QCD axion window in  Eq.~\eqref{Meq},
 $10^8\,{\rm GeV}\lesssim F_a \lesssim 10^{12}\,{\rm GeV}$, one can see that there is an upper bound for masses of axion clumps composed of QCD axion $M_{\rm clump}/M_{\odot}\lesssim 10^{-11}$, where equality is realized for $F_a= 10^{12}\,{\rm GeV}$ and $\alpha = 1$.

Before closing this subsection, we consider the validity of the weak field approximation and nonrelativistic treatment of axion clumps.
To justify the weak field approximation, $R$ should be much longer than the Schwarzschild radius of the clump, $R_S=2G_N M$.  This condition is given by
\begin{align}
    \frac{R}{R_S} &\geq \frac{R_{\rm min}}{2G_N N_{\rm max}m_a} = \frac{\widetilde{R}_{\rm min}}{2\delta \widetilde{N}_{\rm max}}\nonumber \\ 
    &\simeq \frac{0.5\times 10^{-2}}{\delta} \gg 1\,, \label{eq:weak field approximation}
\end{align}
where $\delta \equiv G_N F_a'^2 \simeq 2\times10^{-14}(10^{12}\,\text{GeV}/F_a)^{-2}(0.3/\gamma)$.
Therefore, we see that the weak field approximation holds even for very large axion decay constant.
We next consider the condition of nonrelativistic treatment of axion clumps.
In the original (relativistic) axion potential Eq.~\eqref{eq:axion precisely}, the axion field respects the periodocity $\phi(x) \to \phi(x) + 2\pi F_a$, while the nonrelativistic axion potential does not.
To safely neglect the relativistic corrections to the axion clump configuration, the amplitude of axion field should satisfy the following condition~\cite{Schiappacasse:2017ham}:
\begin{align}
    \frac{\phi_0}{2\pi F_a} = \frac{\Psi_0}{2\pi F \sqrt{2m_a}}=\sqrt{\frac{\delta\widetilde{N}}{56\pi^3\gamma\widetilde{R}^3}} \ll 1 \,,
\end{align}
where $\Psi_0 \equiv \sqrt{N/7\pi R^3}$ is the amplitude of the axion field for the exponential$\times$linear ansatz.
This condition can be reexpressed as 
\begin{align}
    \widetilde{R}&\gg \left(\frac{\delta\widetilde{N}}{56\pi ^3\gamma}\right)^{1/3} \nonumber \\
    &\simeq 4\times 10^{-6}\,\widetilde{N}^{1/3}\left(  \frac{F_a}{10^{12}\,\text{GeV}}\right)^{2/3}\left( \frac{0.3}{\gamma}\right)^{2/3}\,,\label{eq:relativistic approximation}
\end{align}
which is always satisfied for the stable branch because $\delta \ll 1$ and $\tilde{R}$ increases as $\tilde{N}$ decreases. By contrast, for the case of the unstable branch, the size of the clump decreases as the number of particles decreases. The condition in Eq.~(\ref{eq:relativistic approximation}) is no longer justified, and the nonrelativistic approximation breaks down for sufficiently small clump size. In this regime, when the system is analyzed using the relativistic theory, the quasistable branch of axitons emerges as we mentioned before~\cite{Schiappacasse:2017ham, Kolb:1993hw}.

\subsection{Fraction of dark matter in axion stars}
\label{fractionAC}

In this subsection, we give some assumptions to simplify the analysis of the gravitational lensing constraint on the axion clump.

We first assume that axion clumps share the same number of particles $\widetilde{N}$ having a zero-angular momentum. 
This assumption is similar to that of a monochromatic mass function, which is usually assumed in the case of gravitational lensing constraint on PBHs.\footnote{This assumption naturally arises if the axion-photon coupling constant $g_{\phi \gamma \gamma}$ is large enough so that axion clumps may undergo parametric resonance of photons in the early Universe. Since axion clumps can undergo resonance if they have a mass larger than a critical value, we expect today in galactic halos the presence of a pileup of
axion clumps at a unique value of mass. For a detailed discussion, see Secs. 8.1 and 4.2 in Refs.~\cite{Hertzberg:2018zte, Hertzberg:2020dbk}, respectively.}
With this assumption, axion clumps are characterized by four parameters: $m_a$, $F_a$, $\alpha$ and the current fraction of dark matter (DM) in the axion clump, $\Omega_{\rm clump}/\Omega_{\rm DM}$. 
For the axion clump formed by the ordinary QCD axion, the axion mass and its decay constant are related. The fraction of axion clumps in DM depends on the scenario of their formation.
In the standard postinflationary scenario,
$\Omega_{\rm clump}/\Omega_{\rm DM}\simeq 1/10$ ~\cite{Guth:2014hsa,Schiappacasse:2017ham}.
When the PQ symmetry breaking takes place after the inflation, the correlation length of the axion before the QCD phase transition is the order of particle horizon $1/H (T_{\rm QCD})$. Here, $H (T_{\rm QCD})$ is the Hubble parameter at the temperature of the QCD phase transition, $T_{\rm QCD}$.

Assuming that the axion constitutes the whole DM density, we can estimate the axion number density at $T_{\rm QCD}$ as
\begin{align}
n(T_{\rm QCD}) = \frac{\rho_{\rm DM}(T_{\rm QCD})}{m_a} &=  \frac{T_{\rm eq}}{T_{\rm QCD}}  \frac{\rho_{\rm rad} (T_{\rm QCD})}{m_a} \nonumber\\
 &\sim \frac{T_{\rm eq} T_{\rm QCD}^3}{m_a},
\end{align}
where $T_{\rm eq}\sim 0.1{\rm eV}$ is the temperature at matter-radiation equality, $\rho_{\text{DM}}$ is the DM density, and $\rho_{\text{rad}}(T)$ is the radiation density at temperature $T$.
Within the correlation length $1/H (T_{\rm QCD})$, the total number of axions is therefore estimated as 
\begin{align}
N \sim \frac{n(T_{\rm QCD})}{H^3(T_{\rm QCD})} \sim \frac{T_{\rm eq}M^3_{\rm Pl}}{T_{\rm QCD}^3m_a} \sim 10\times N^{\rm QCD}_{\rm max},
\end{align}
where $N^{\rm QCD}_{\rm max} \sim 10^{60}$ [see Eqs. \eqref{eq:particle numbers}, \eqref{eq:Nmax}, and \eqref{Neq}] is the maximum number of ordinary QCD axions within the axion clump. Thus, we may expect that the current dark matter fraction in axion clumps is about  $10\%$ or less \footnote{Here we are only considering spherically symmetric clumps. Including a non-zero angular momentum leads to a maximum number of particles which depends on the clump angular configuration. In particular, for quantum numbers $l = |m| = 5$, we have $N_{\text{max}}^{\text{QCD}} \gtrsim 10^{61}$. This number now is large enough to  accommodate all the available axions in a typical correlation length (for further details, see Secs. 3 and 6 in Ref.~\cite{Hertzberg:2018lmt}).}.

On the other hand, in the preinflationary scenario, when the PQ symmetry breaking takes place before or during inflation, the axion field becomes highly homogeneous on large scales. Thus, it is unclear that axions may form a BEC in the late Universe in the standard way that we explained above. However, as we mentioned in the Introduction, the kinetic nucleation of QCD or string axion clumps in minihalos around
PBHs~\cite{Hertzberg:2020hsz} and axionlike particles clumps formed via tachyonic instability driven by a multiple cosine potential~\cite{Fukunaga:2020mvq} can occur for general values of the PQ scale. In these cases, the estimation of the fraction of dark matter in axion clumps needs numerical simulations. For example, for the case of axion clumps nucleation around PBHs, the final fraction of axion clumps in DM significantly depends on the factors such as the fraction of PBHs and the number of clumps nucleated by minihalos.

Keeping in mind the above discussion, we leave the fraction $\Omega_{\text{clumps}}/\Omega_{\text{DM}}$ as a free parameter when we discuss gravitational lensing constraints in the next sections.

\section{Microlensing by axion stars}\label{sec:microlensing}

In this section, we briefly review basics of gravitational microlensing events~\cite{1986ApJ...304....1P}.
In particular, we calculate the threshold impact parameter for a spherically symmetric axion clump configuration by solving the lens equation including the finite lens and source size effects.
Microlensing constraints on other compact objects such as boson stars and axion miniclusters were investigated in Refs.~\cite{Kolb:1995bu,Fairbairn:2017sil,Croon:2020wpr,Croon:2020ouk,Marfatia:2021twj}.

This section is organized as follows.
In Sec.~\ref{sec:point lens point source}, we estimate a magnification caused by a pointlike lens and calculate the threshold impact parameter.
In Sec.~\ref{sec:non-point lens}, we take into account a finite lens size effect for the calculation of the threshold impact parameter.
In Sec.~\ref{sec:finite size source effect}, the finite source size effect is dealt with in addition to the finite lens size effect.
A detailed derivation of the lens equation is summarized in the Appendix.~\ref{app:appendixA}.
A theoretical estimate of the expected number of microlensing events with a given threshold impact parameter is discussed in Sec.~\ref{sec:event rate and microlensing constraints}.

Before going to the detailed analysis, let us give a basic setup of a microlensing event.
We mainly follow the treatment discussed in Ref.~\cite{Narayan:1996ba}.
\begin{figure*}[!t]
\begin{center}
\includegraphics[clip, width=10cm]{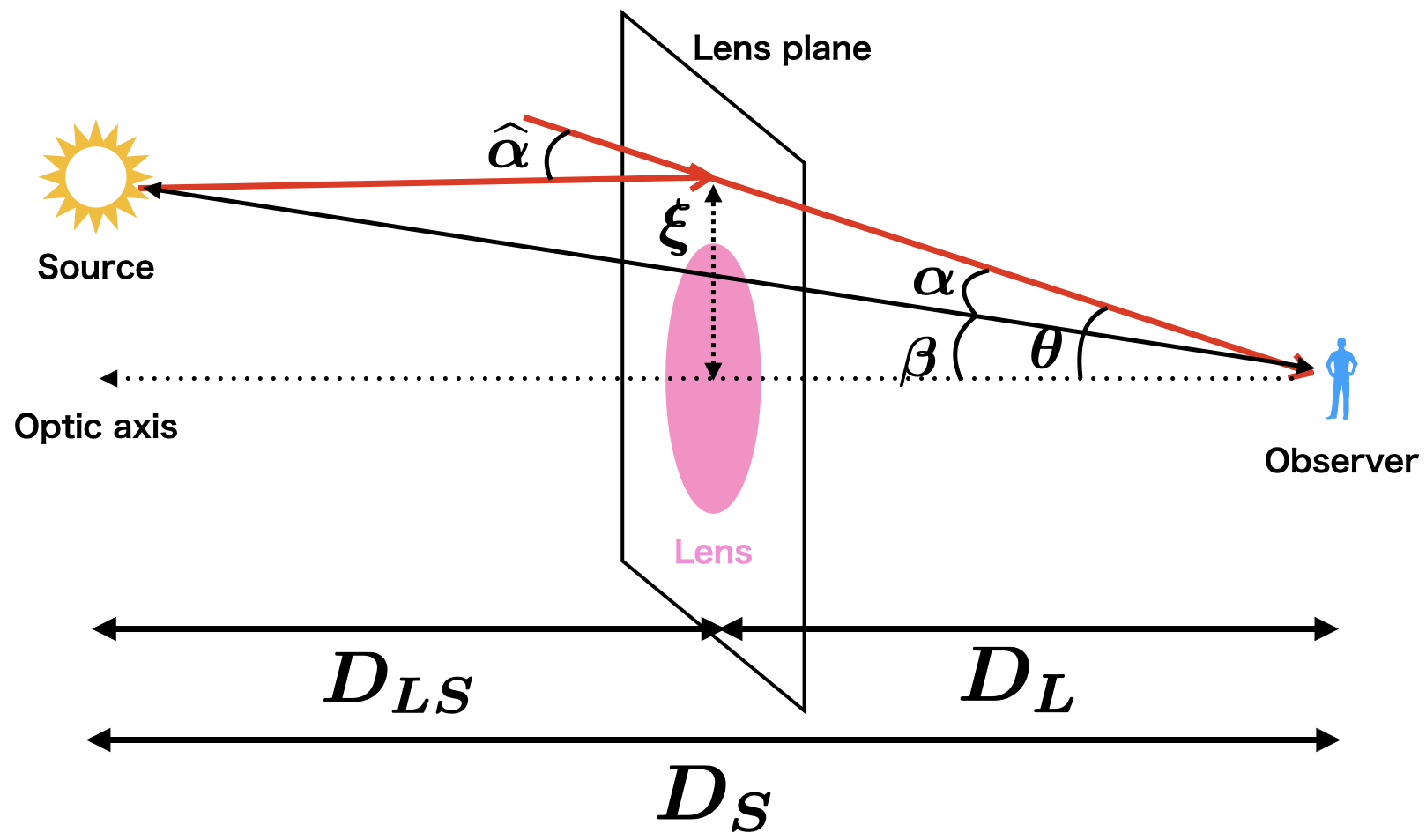}
\end{center}
\caption{
A geometrical setup of the gravitational lensing event is shown.
A light ray (the red color line) is deflected by the lens (the pink colored circle) and reaches to the observer (the blue colored human).
}\label{fig:microlensing setup}
\end{figure*}
A geometrical setup of a microlensing event is shown in Fig.~\ref{fig:microlensing setup}.
In the figure, we take an optic axis in such a way that an observer and the center of the lens (the axion clump) are aligned with each other assuming that a source star is a pointlike (although we consider finite source size effect in Sec.~\ref{sec:finite size source effect}).
Mass distribution of the lens (the pink colored circle) is projected onto the lens plane, which is taken to be orthogonal to the line of sight.
A light ray emitted by the source star is deflected with the angle $\widehat{\alpha}$ at the lens plane and reaches to the observer.
The diameter distances from the observer to the source, to the lens, and from the lens to the source are
$D_{\rm S},~D_{\rm L}$ and $D_{\rm LS}$, respectively.
The reduced angle of $\widehat{\alpha}$, $\alpha$, is explicitly estimated in the Appendix.
The angle between the optic axis and the line from the observer to the true position of the source, and that between the optic axis and the line from the observer to an image of the source, are denoted by $\beta$ and $\theta$, respectively.
Note that multiple images $\theta_i$ ($i=1,2,\cdots$), corresponding to the single source position $\beta$, are generally observed, but we only show one example in the figure for simplicity.

\subsection{Microlensing by a point lens with a point source}\label{sec:point lens point source}

With the setup shown in Fig.~\ref{fig:microlensing setup}, a lens equation with a point lens and a point source is given by~\cite{Narayan:1996ba}
\begin{align}
    \beta = \theta-\frac{D_{\rm LS}}{D_{\rm L}D_{\rm S}}\frac{4G_NM_{\rm clump}}{\theta}.\label{eq:lens equation point mass lens}
\end{align}
The derivation of the above equation is shown in the Appendix.
The pointlike Einstein ring angle $\theta_E$ is defined as a solution of the above equation with $\beta(\theta_E)\equiv 0$,
\begin{align}
    \theta_E = \sqrt{4G_NM_{\rm clump} \frac{D_{\rm LS}}{D_{\rm S}D_{\rm L}}}.
\end{align}
The pointlike Einstein ring radius on the lens plane is given by $R_E\equiv D_L\theta_E$.
The typical value of the Einstein ring radius is estimated as
\begin{align}
	R_E \simeq 1.9\times 10^3&\times R_{\odot}\nonumber \\
	&\times \left( \frac{M_{\rm clump}}{M_{\odot}}\right)^{\frac{1}{2}}\times\left(\frac{D_{\rm S}}{10\,{\rm kpc}}\right)^{\frac{1}{2}}, \label{eq:Einstein ring radius typical value}
\end{align}
where $R_{\odot}\simeq 7.0\times 10^8\,{\rm m}$ is the solar radius.
In this calculation, we have used $D_{\rm S}\sim D_{\rm L}\sim D_{\rm LS}$, which gives a good order estimation.
Then the lens equation, Eq.~\eqref{eq:lens equation point mass lens}, takes a simple form and has the following two solutions:
\begin{align}
&u = t -\frac{1}{t}\,,\nonumber \\
&t_{1} = \frac{u}{2} \left(1 + \sqrt{1 +\frac{4}{u^2}}\right)\,,\label{eq:solutions of lens equation}\\
&t_{2} = \frac{u}{2} \left(1 - \sqrt{1 +\frac{4}{u^2}}\right)\,, \nonumber
\end{align}
where $u=\beta/\theta_E$ and $t_i = \theta_i/\theta_E$.

The gravitational lensing does not change the surface brightness of the source star, but does change the apparent area of the source image because the observer receives a total flux magnified by the lens.
Throughout this paper, we assume that the coupling between the lens (the axion clump) and a photon (a light ray) is not so large to distort the image of the source, that is, the lensing event is purely caused by the gravitational effect.\footnote{A lensing event induced by the photon-axion coupling was investigated in Ref.~\cite{Prabhu:2020pzm}.}
For a cylindrically symmetric lens, a magnification $\mu_i$ caused by the source image $\theta_i$ is defined by the ratio of an image area to a source area.
For a pointlike source, it is given by the ratio of the solid angle of the image area to that of the source area,
\begin{align}
\mu_i \equiv \frac{\theta_i}{\beta}\frac{d\theta_i}{d\beta} = \frac{t_i}{u}\frac{dt_i}{du},\label{eq:magnification with point source}
\end{align}
where $\theta_i~(i=1,2)$ is the solution of the lens equation, Eq.~\eqref{eq:solutions of lens equation}.
A total magnification of two images is given by
\begin{align}
\mu_{\rm tot} (u)=|\mu_1| + |\mu_2| = \frac{u^2+2}{u\sqrt{u^2+4}},~u\equiv \beta/\theta_E. \label{eq:magnification by point mass lens}
\end{align}

The EROS-2 survey and Subaru HSC observation use microlensing event selections through the criterion that magnifications of source stars exceed the threshold value $\mu_T = 1.34$.
According to this criterion, a microlensing event therefore occurs when $\mu >1.34$ is realized.
A threshold impact parameter, $u_T$, is defined by 
\begin{align}
	\mu_{\rm tot}(u = u_T) = 1.34.
\end{align}
This implies that impact parameters smaller than this threshold value cause a microlensing event.
Since we assume a point lens and a point source, the threshold impact parameter is given by $u_T = 1$, which is completely the same as that of PBHs with a point source.
As we see later, the value of threshold impact parameter becomes different from this value when we take account of the finite lens and the finite source effects.

\subsection{A finite lens size effect}\label{sec:non-point lens}

The main purpose of this subsection is to calculate the threshold impact parameter including the finite size lens effect, which was neglected in the previous subsection.

The lens equation including the extent of the axion clump is derived in the Appendix and is given by
\begin{align}
\beta (\theta) = \theta - \frac{D_{\rm LS}}{D_{\rm L} D_{\rm S}} \frac{4G\mathcal{M}(\xi_{\rm L})}{\theta}, \label{eq:lens equation}
\end{align}
where $\mathcal{M}$ is the total mass of the axion clump projected onto the lens plane defined in \eqref{eq:mass projected onto lens plane} and $\xi_{\rm L}\equiv D_{\rm L}\theta$.
For the axion field configuration given by Eq.~\eqref{eq:ansatz}, $\mathcal{M}(\xi)$ can be expressed as
\begin{align}
    \mathcal{M}(\xi)=MG(w),~w\equiv \xi/R,
\end{align}
\begin{align}
    G(w)&\equiv\frac{2}{7} \int^{\infty}_{-\infty}dz'\nonumber \\
    &\times \int^w_0 dw'w'\left(1+\sqrt{w'^2+z'^2}\right)^2e^{-2\sqrt{z'^2+w'^2}}.
\end{align}
Here, $\mathcal{M}$ and $G(w)$ represent the effective mass of the axion clump within the radius $\xi$ and the axion clump configuration projected onto a lens plane, respectively.
The lens equation including the extent of the axion clump becomes
\begin{align}
    u = t-\frac{G(w_E t)}{t},\label{eq:lens eq}
\end{align}
where $w_E\equiv D_L \theta_E= R_E /R$.
In this expression, the function $G(w_E t)$ represents the finite size lens effect parametrized by the parameter $w_E$.

We now qualitatively discuss the finite lens size effect.
For large $w_E\gg 1$, since the pointlike Einstein ring radius, which is the characteristic length scale of the microlensing event, is much longer than the characteristic length scale of the axion clump, the finite size lens effect becomes unimportant.
Indeed, $w_E \to \infty$ with fixed $t$ gives $G(w_E t)\to 1$, which recovers the lens equation for a pointlike lens given by Eq.~\eqref{eq:solutions of lens equation}.
On the other hand, in the opposite case $w_E \ll 1$, the extent of the axion clump becomes important.
In this regime, since the mass of the axion clump within the pointlike Einstein ring radius is small compared to the point lens one, the total magnification is significantly suppressed by the finite lens size effect.
Indeed, $w_E\to 0$ with fixed $t$ gives $G(w_E t )\to 0$, and hence, one immediately obtains $u=t$ corresponding to no magnification $\mu_{\rm tot}=1$.
Thus, there is no microlensing constraint in this regime.

To quantitatively discuss the finite lens size effect, let us exactly solve the lens equation, Eq.~\eqref{eq:lens eq}, and compute a magnification with given $w_E$.
Unfortunately, $G(w_E t)$ is a highly nonlinear function, and thus, one cannot analytically solve Eq.~\eqref{eq:lens eq}.\footnote{Even if one fits the $G(w)$ function with polynomial functions, the lens equation becomes a polynomial equation higher than a quintic equation, which cannot be analytically solved.
}
By numerically solving this equation with given $u$ and $w_E$, one obtains multiple solutions $t_i=t_{i}(u,w_E)$.
From the definition of the magnification by a point source given by Eq.~\eqref{eq:magnification with point source}, a magnification can be expressed in terms of $u$ and $w_E$ through $w_{{\rm L}_i}\equiv w_E t_i$ as
\begin{align}
\mu_i(u,w_E)=\frac{1}{(1-B_i)(1+B_i-C_i)},
\end{align}
where
\begin{align}
   B_i=\frac{w_E^2}{w_{{\rm L}_{i}}^2}G(w_{{\rm L}_{i}}),~C_i =\frac{w_E^2}{w_{{\rm L}_i}}G'(w_{{\rm L}_i}). \label{eq:C}
\end{align}
In this expression, the prime denotes the derivative with respect to $w_{{\rm L}_i}$.
The total magnification is given by the sum of absolute values of $\mu_i$, $\mu_{\rm tot}=\sum_i |\mu_i|$.
Imposing $\mu_{\rm tot}(u=u_T) = 1.34$, one can evaluate the threshold impact parameter $u_T(w_E)$ including the extent of the axion clump.
Note that $C_i$ in the above expression represents a gradient contribution of the lens, which is absent in the pointlike lens limit.
This gives an additional contribution to the magnification, which is discussed later.

\begin{figure*}[!t]
\begin{center}
\includegraphics[clip, width=5.cm]{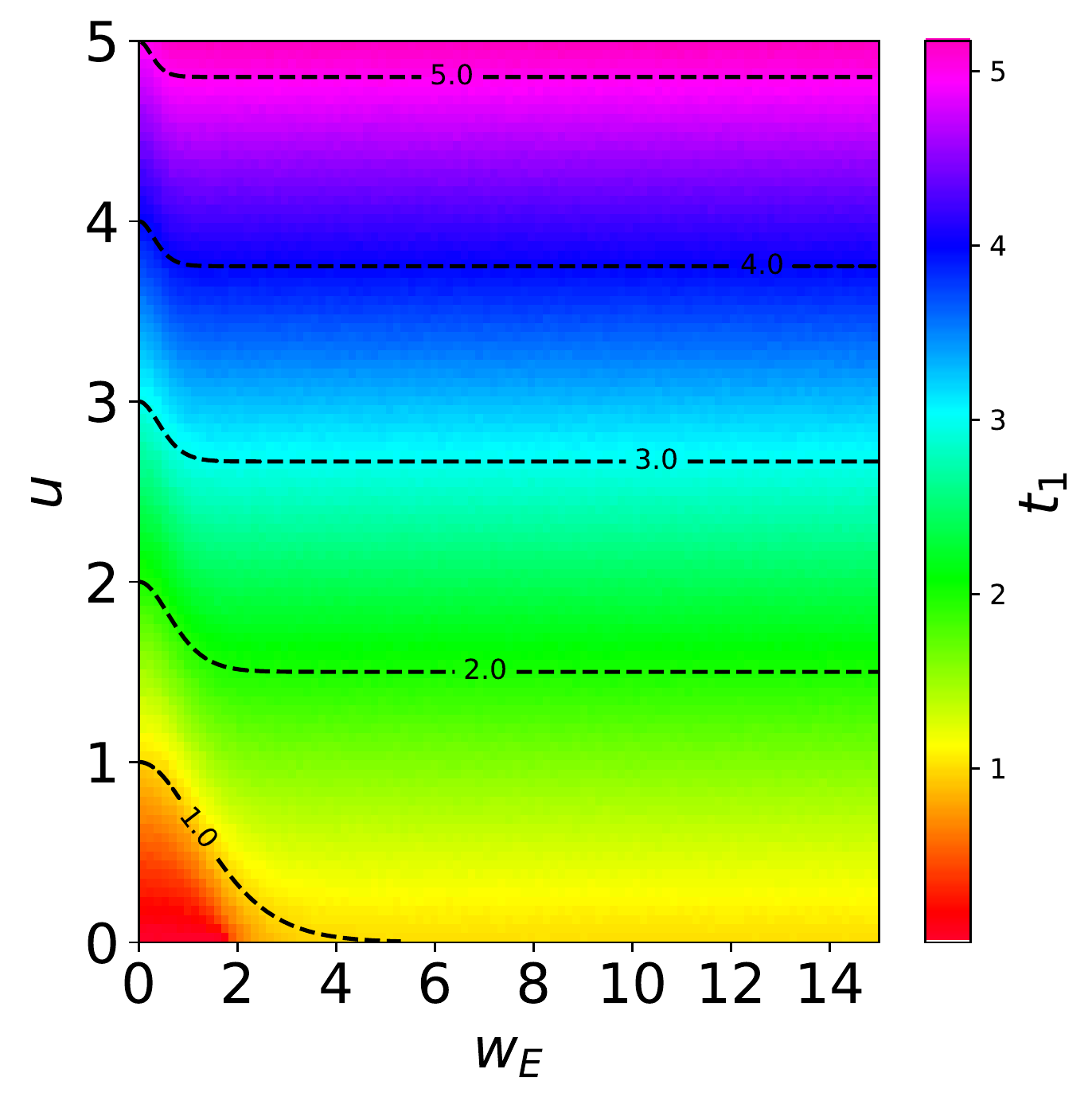}
\includegraphics[clip, width=5.cm]{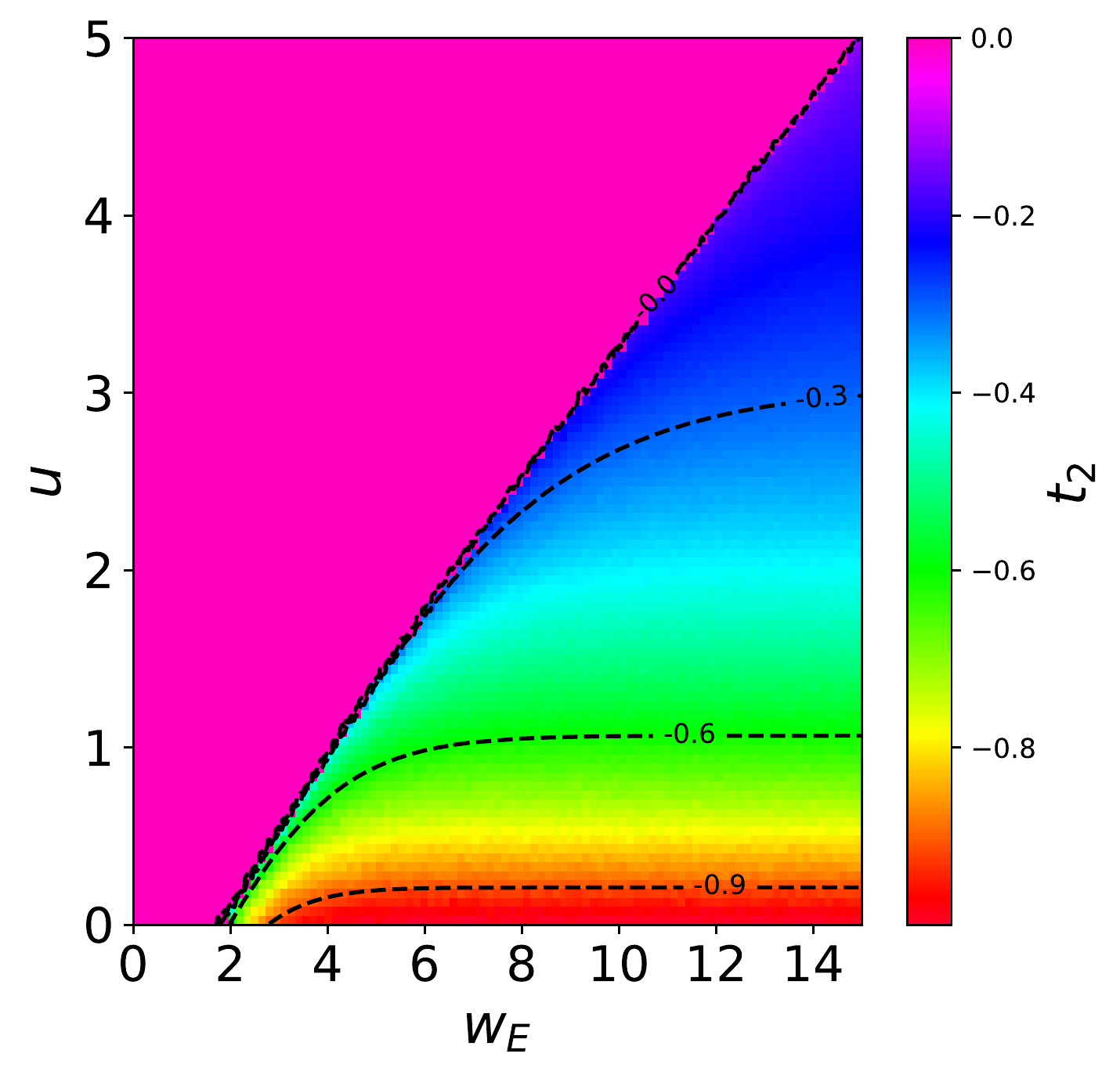}
\includegraphics[clip, width=5.cm]{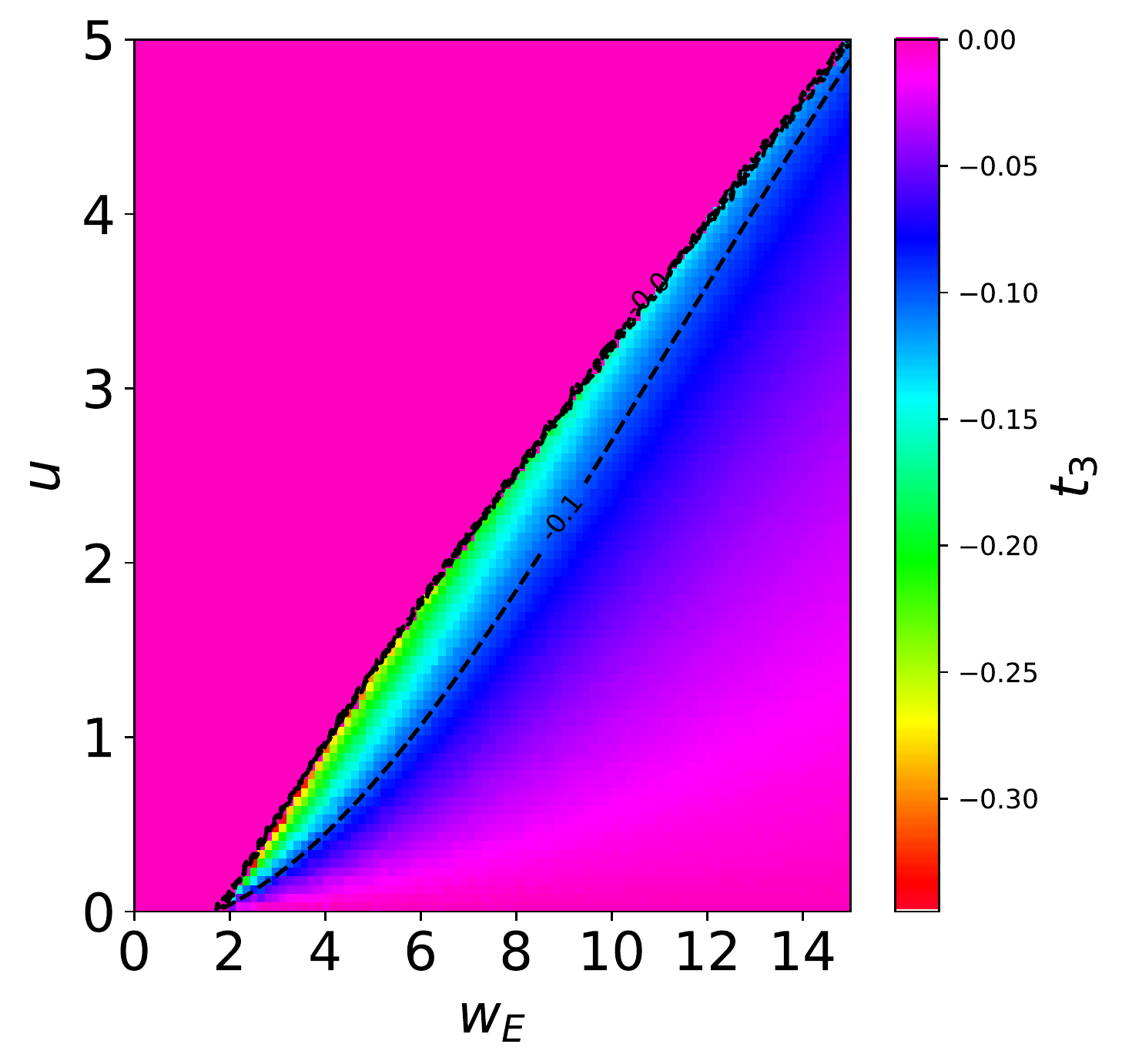}
\end{center}
\caption{Solutions of the lens equation, Eq.~\eqref{eq:lens equation}: $t_1$ (left), $t_2$ (middle) and $t_3$ (right) are shown as a color code for the axion field configuration approximated by an exponential$\times$linear ansatz.
In each panel, the black dashed contour curves for $t_i$ are shown.
\label{fig:solutions of lens equation}
}
\end{figure*}
We show solutions of the lens equation, Eq.~\eqref{eq:lens eq}, in Fig.~\ref{fig:solutions of lens equation}.
In the figure, the axion field configuration is approximated by the exponential$\times$linear ansatz given by Eq.~\eqref{eq:ansatz}.
The number of solutions are maximally three, $t_{1,2,3}$, depending on the values of $w_E$ and $u$.
At $w_E \to \infty$ with fixed $u$, $t_{1,2}$ solutions become those given by Eq.~\eqref{eq:solutions of lens equation}, leading to the magnification by a pointlike  lens, Eq.~\eqref{eq:magnification by point mass lens}.
At the same time, $t_3$ solution is nonzero, but it is vanishingly small $t_3\simeq 0$, which leads to a negligible magnification $\mu_3\simeq 0$.
Thus, $w_E\to \infty$ can be regarded as a pointlike lens limit.
When we make $w_E$ small with fixed $u$, $t_{2}$ and $t_{3}$ solutions eventually coincide with each other, $t_2= t_3$, and these solutions $t_2$ and $t_3$ discontinuously vanish, $t_{2,3}=0$
(see the middle and right panels of Fig.~\ref{fig:solutions of lens equation}).
 For smaller $w_E$, a single image $t_1\neq 0$ still remains. 
Magnifications caused by these sudden change of  degenerated $t_{2,3}$ are strongly enhanced because $du / dt \simeq 0$ is realized.\footnote{However, the magnification never diverges because it is regulated by the finite source size.}
In the limit of $w_E\to 0$, one obtains $t_1 \to u$ leading to a no magnification $\mu_1=1$ due to the significant finite lens size effect.
We confirm that these behaviors are in agreement with the results obtained in Ref.~\cite{Croon:2020wpr} for a spherical lens with uniform density.

\begin{figure}[!t]
\begin{center}
\includegraphics[clip, width=7cm]{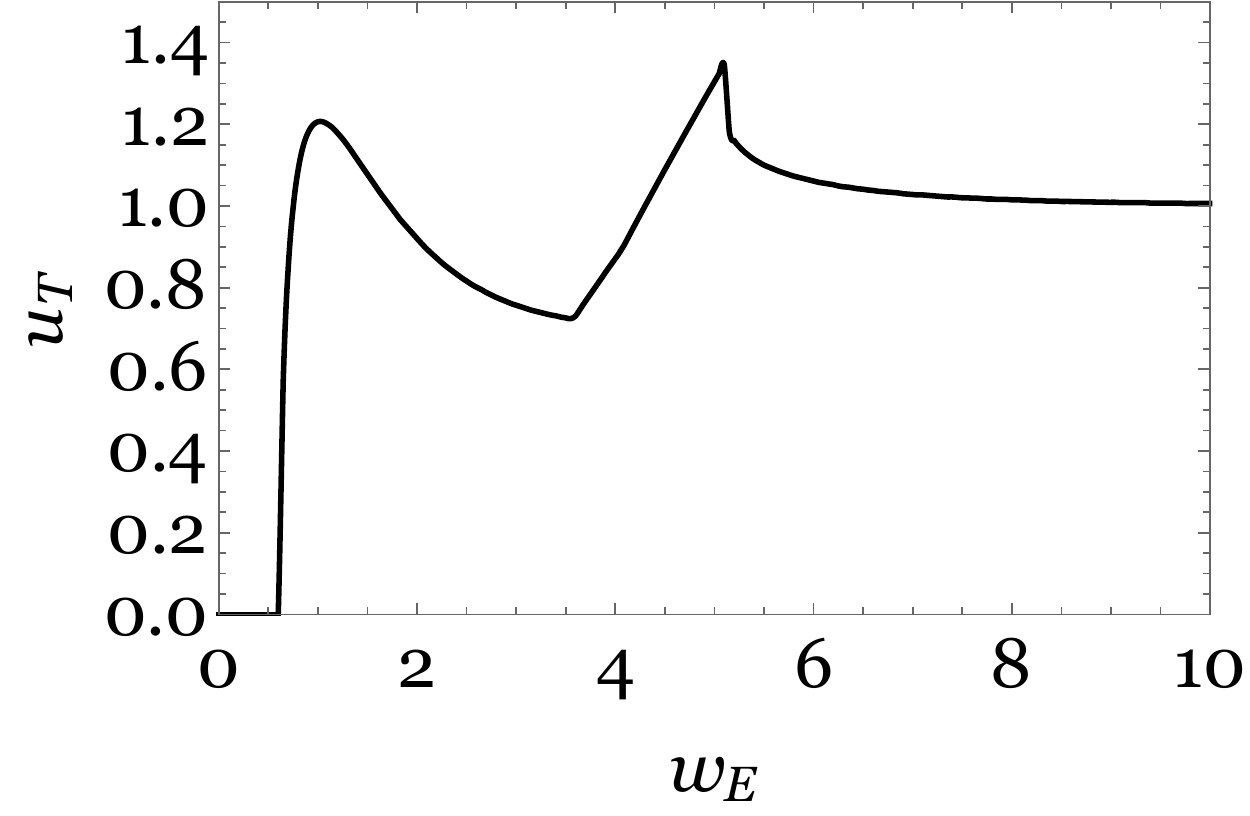}
\end{center}
\caption{Threshold impact parameters with a linear$\times$exponential ansatz is shown as a function of $w_E\equiv R_E/R$.
}\label{fig:impact parameter}
\end{figure}

The threshold impact parameters evaluated by using the linear $\times$ exponential ansatz are shown in Fig.~\ref{fig:impact parameter}.
As is expected, the point lens result with $u_T=1$ can be obtained for a large $w_E$.
We confirm that the axion clump can be identified with a pointlike lens for $w_E\gtrsim 7$ within 1\% accuracy.
For intermediate regime $0.6\lesssim w_E\lesssim 7$, a finite lens size correction is non-negligible and we can observe interesting behavior of $u_T$.
As we noted in the previous paragraph, a magnification caused by degenerated $t_{2,3}$ is enhanced, and thus, $u_T > 1$ can be realized at around $w_E \simeq 5$.
For a smaller $w_E $, a magnification is only sourced by an image $t_+$ but it is also slightly enhanced, $u_T > 1$, by the gradient term $C_i$ in Eq.~\eqref{eq:C}.
For very small $w_E\lesssim0.6$, one obtains $u_T=0$ due to a significant suppression from the finite lens size effect, and thus, there is no microlensing constraint in this parameter region.
We confirm that this behavior is in agreement with that for boson stars shown by Fig.~2 in Ref.~\cite{Croon:2020wpr}.

We have seen that the threshold impact parameter including the finite lens size effect is parametrized by $w_E$.
It is helpful to express the $w_E$ parameter in terms of fundamental parameters such as $m_a,~F_a$, and $\alpha$ as follows:
\begin{align}
	w_E \simeq 10^2&\times \left( \frac{m_a}{10^{-5}\,{\rm eV}}\right)^{\frac{1}{2}} \left( \frac{F_a}{10^{12}\,{\rm GeV}}\right)^{\frac{3}{2}} \nonumber \\
	&\times \left( \frac{\gamma}{0.3}\right)^{\frac{3}{4}}\left(\frac{D_{\rm S}}{10\,{\rm kpc}}\right)^{\frac{1}{2}}\left(\frac{\alpha^{\frac{3}{2}}}{1+\sqrt{1-\alpha^2}}\right). \label{eq:wE}
\end{align}
In this calculation, we have taken $D_{\rm S} \sim D_{\rm LS} \sim D_L$, which gives a good order estimation.
This order estimate helps us to understand the behavior of the microlensing constraint including the finite lens size effect.

\subsection{Finite lens size and finite source size effects}\label{sec:finite size source effect}

So far, we have estimated the threshold impact parameter assuming that source stars are pointlike, which is only valid when source star radii are much smaller than the pointlike Einstein ring radius.
We follow the method described in Ref.~\cite{1994ApJ...430..505W,Montero-Camacho:2019jte,Croon:2020ouk} to include the finite source size effect in this subsection.

\begin{figure}[!t]
\begin{center}
\includegraphics[clip, width=7.cm]{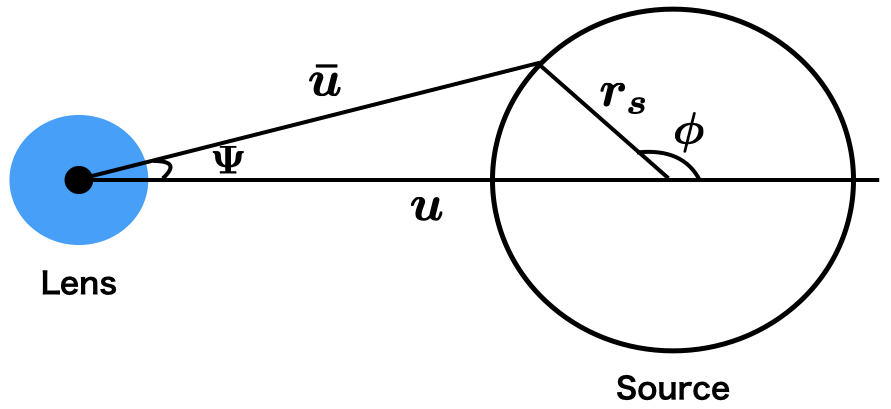}
\end{center}
\caption{
A source star (the black circle) and the lens (the blue colored blob circle) projected onto the lens plane are shown in the figure.
 All distances are normalized by the Einstein ring radius.
\label{fig:finite source size setup}
}
\end{figure}
Since radii of source stars are, typically, many orders of magnitude shorter than $D_{\rm S},D_{\rm L}$, and $D_{\rm LS}$, we only need to consider the extent of the source projected onto a lens plane.
For simplicity, we assume that a source star is spherically symmetric.
We show our setup in Fig.~\ref{fig:finite source size setup}.
In the figure, all distances are normalized by the pointlike Einstein ring radius.
$u$ and $r_s$ are the impact parameters from the center of the source star and a source radius on the lens plane defined as $r_s \equiv x R_{\rm S} / R_E$, respectively, where $x\equiv D_{\rm L}/D_{\rm S}$, and $R_{\rm S}$ is the source star radius.

An impact parameter on the edge of the source $\overline{u}(u,\phi,w_s)$ can be expressed as follows:
\begin{align}
    \overline{u}(u,\phi,r_s)=\sqrt{u^2+r_s^2+2u r_s\cos\phi}.\label{eq:impact parameter on the edge of the source}
\end{align}
Then a lens equation for $\bar{u}(u,\phi,r_s)$ becomes
\begin{align}
    \overline{u}(u,\phi,r_s)=t-\frac{G(w_E t)}{t}.\label{eq:lens equation finite size source}
\end{align}
By numerically solving the above equation with given $u,w_E,\phi$, and $r_s$, one obtains multiple solutions $t_i = t_i(u,w_E,\phi,r_s)$.
As was seen in the previous subsection, the number of solutions of the above equation is again maximally three, $t_{1,2,3}$, which are shown in Fig.~\ref{fig:solutions of lens equation} with a replacement of $u\to\overline{u}(u,\phi,r_s)$.
The magnification caused by the image $t_i$ is defined by the ratio of the image area to the source area~\cite{1994ApJ...430..505W,Montero-Camacho:2019jte},
\begin{align}
    &\mu_i (u,w_E,r_s)=\frac{(-)^{P_i}}{2\pi r_s^2} \oint d\Psi  t_i^2,\nonumber \\
    &\tan\Psi\equiv \dfrac{r_s\sin\phi}{u+r_s\cos\phi},\label{eq:magnification finite source size}
\end{align}
where $(-)^{P_i}$ represents the parity factor of the image and the integration is taken over the edge of the source corresponding to $\phi=0$ to $\phi=2\pi$.
Here, $(-)^{P_i}=+1$ ($i=1,3$), and $(-)^{P_i}=-1$ ($i=2$).
The total magnification is then given by $\mu_{\rm tot}(u,w_E,r_s)=\sum^3_{i=1} \mu_i$.
The threshold impact parameter is thus estimated by imposing the condition $\mu_{\rm tot}(u=u_T,w_E,r_s)=1.34$ for fixed $w_E$ and $r_s$.

Let us here qualitatively discuss the finite source size and the finite lens size effects.
From Eq.~\eqref{eq:impact parameter on the edge of the source}, it is obvious that the impact parameter on the edge of the source is bounded below $\bar{u}(u,\phi,r_s)\gtrsim r_s$ for $u\ll r_s$.
This reflects the fact that the impact parameter cannot be zero due to the finite source size effect.
In the limit of $r_s\to \infty$ with fixed $w_E$, one immediately obtains $\bar{u}=t_1$ corresponding to no magnification.
Thus, the threshold impact parameter is significantly suppressed by the finite source size effect for $r_s \gtrsim 1$.

\begin{figure}[!t]
\begin{center}
\includegraphics[clip, width=8.cm]{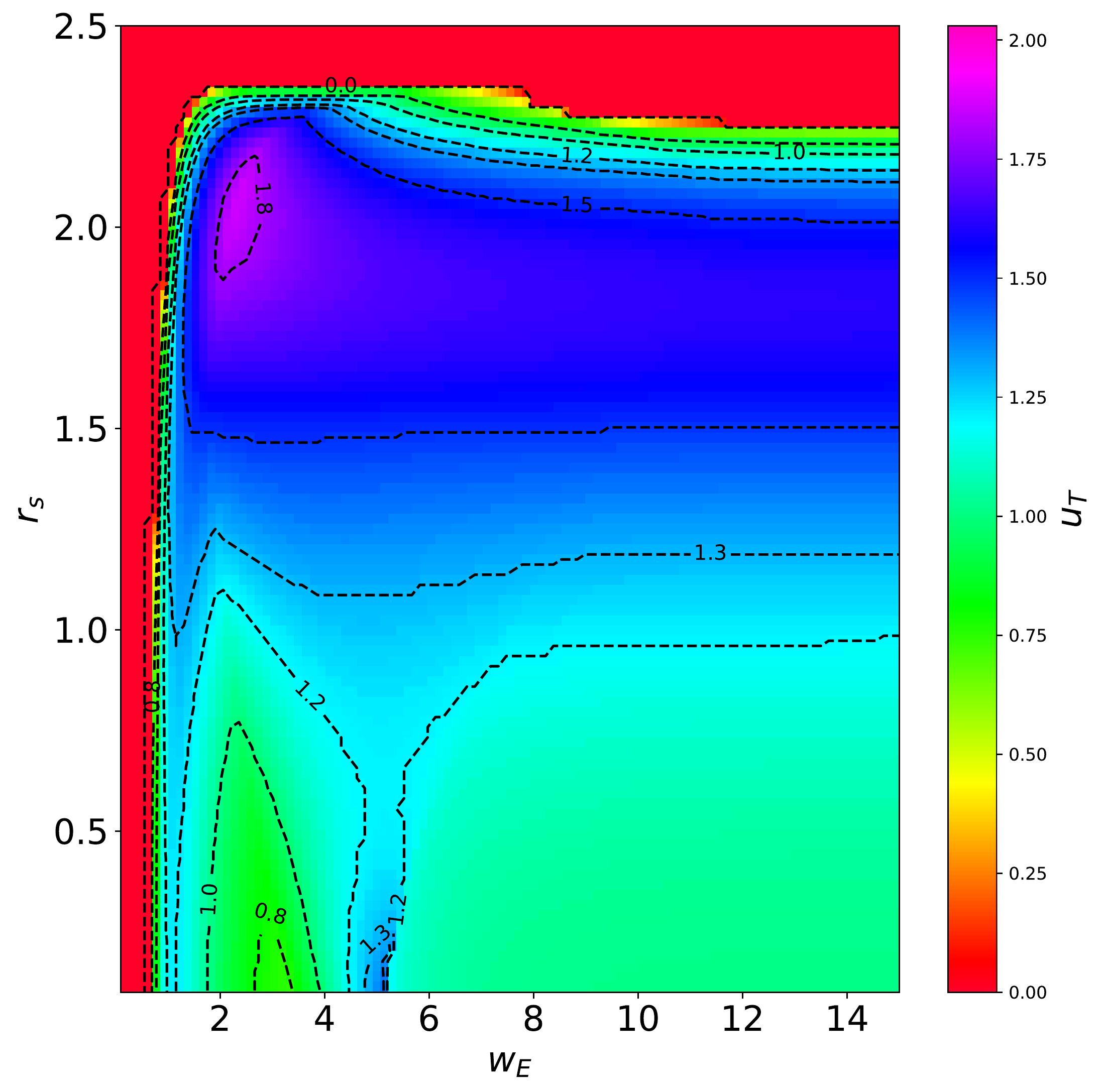}
\end{center}
\caption{
A threshold impact parameter $u_T$ on the $(w_E,r_s)$ plane is shown as a color code combined with black dashed contour curves for $u_T=0,~0.8,~1.0~,1.2,~1.5,~1.8$.
\label{fig:finite_size_source}
}
\end{figure}
We now quantitatively discuss the finite source size effect.
The threshold impact parameter including the finite source and lens size effects is shown in Fig.~\ref{fig:finite_size_source} as a color code.
We confirm that a large source radius $r_s \gtrsim 2.3$ with arbitrary $w_E$ gives $u_T=0$ due to the suppression from a finite source size effect.
Furthermore, for $w_E\lesssim 0.6$, one obtains $u_T=0$ with arbitrary $r_s$ due to the suppression from the finite lens size effect, which was discussed in the previous subsection.
Hence these parameter regions cannot be constrained by observations of microlensing events.
Note that, in the limit $r_s \to 0$ with $w_E\neq 0$, the pointlike source result shown in Fig.~\ref{fig:impact parameter} is reproduced.
Moreover, in the limit $w_E\to \infty$ with $r_s\neq 0$, the pointlike lens result, which was obtained in the context of PBHs~\cite{Smyth:2019whb}, is also reproduced.

\section{Event rates and microlensing constraints}\label{sec:event rate and microlensing constraints}

In the previous section, we have calculated the threshold impact parameter including the finite lens and source size effects, which must be included to give microlensing constraints on axion clumps.
In this section, we estimate the expected number of microlensing events with given $u_T$.
In particular, we focus on the EROS-2 survey~\cite{Tisserand:2006zx} as well as the Subaru HSC observation~\cite{Niikura:2017zjd}.

A differential event rate per unit source star and per unit time with given $u_T$ was estimated
in Refs.~\cite{Griest:1990vu,Alcock:1995zx} for generic compact objects where a velocity distribution of the compact object is assumed to be a Maxwell-Boltzmann distribution.
It can be expressed as
\small
\begin{align}
\frac{d\Gamma}{d\hat{t}} = \epsilon(\hat{t})2D_{\rm S}\frac{\Omega_{\rm clump}}{\Omega_{\rm DM}}\int^{1}_0 \frac{\rho_{\rm DM}(x)}{M} Q(x)^2 v_c^2 e^{-Q(x)} dx. \label{eq:differential event rate}
\end{align}
\normalsize
where $Q(x)\equiv 4R_E^2(x)u_T^2(x) / \hat{t}^2 v_c^2$.
In this expression, $v_c,~\hat{t},~\epsilon(\hat{t})$, and $\rho_{\rm DM}$ are the dark matter circular velocity in the galaxy, the time to cross the {\it Einstein ring diameter}, the efficiency factor, and the dark matter density in the halo, respectively.
The expected number of events $N_{\rm exp}$ is then estimated as
\begin{align}
N_{\rm exp} = E \int^\infty_0 \frac{d\Gamma}{d\hat{t}}d\hat{t},\label{eq:expected number of events}
\end{align}
where $E$ is the exposure time in sidereal years.
We also assume that the number of microlensing events follows the Poisson distribution.
Under this assumption, the probability to observe $N_{\rm obs}$ numbers of microlensing events with a given $N_{\rm exp}$ can be estimated as $P(N_{\rm obs},N_{\rm exp})=(N_{\rm exp})^{N_{\rm obs}}e^{-N_{\rm exp}}/N_{\rm obs}!$\,.
Therefore, one can exclude the parameter region leading to $\sum_{k=0}^{N_{\rm obs}} P(k,N_{\rm exp})<0.05$ with given $N_{\rm obs}$ corresponding to a $95\%$ confidence level.
In order to calculate $N_{\rm exp}$, the variables $D_{\rm S},~\epsilon(\hat{t}),~\rho_{\rm DM}(x),~E,~v_c$ and $N_{\rm obs}$ need to be specified, but these variables highly depend on which observation we used.
Hence, we explicitly clarify the setups of the EROS-2 survey and the Subaru HSC observation in the following subsections.

\subsection{The EROS-2 survey}\label{sec:EROS-2}

In this subsection, we show microlensing constraints on the axion clump by using the EROS-2 survey.

Let us first clarify the setup of the EROS-2 survey.
The EROS-2 survey focuses on the source stars in the large Magellanic cloud (LMC), whose distance is given by $D_S \simeq 50 {\rm kpc}$, and the circular velocity for the MW is approximately given by $v_c\simeq 220\,{\rm km/s}$~\cite{2019ApJ...871..120E}.
We do not include the small Magellanic cloud data in our analysis since its effect is subdominant and does not change our conclusion significantly.
We here assume an isothermal profile for the MW halo~\cite{Cirelli:2010xx,Croon:2020wpr},
\begin{align}
&\rho_{\rm DM} (D_L) = \dfrac{\rho_s}{1+(r/r_{\rm iso})^2},\nonumber\\
&r^2  (D_L)= R_{\rm sol}^2-2R_{\rm sol}D_L\cos (l) \cos (b)+D_L^2, 
\end{align}
where $\rho_s=1.39\,{\rm GeV/cm^3},r_{\rm iso}=4.38\,{\rm kpc},~R_{\rm sol}=8.5\,{\rm kpc}$ and $(l,b)=(\ang{280},-\ang{33})$, respectively.
The exposure time is $E\simeq 3.77\times 10^7$ sidereal years.
The efficiency factor, $\epsilon(\hat{t})$, is taken from Fig.~11 shown in Ref.~\cite{Tisserand:2006zx}.
Note that the efficiency factor is given in terms of the time to cross the {\it Einstein ring radius}, $t_E=\widehat{t}/2$.
Under this setup, one can evaluate $N_{\rm exp}$ defined by Eq.~\eqref{eq:expected number of events} with fixed $m_a$, $\alpha$, and $\Omega_{\rm clump}/\Omega_{\rm DM}$ for the ordinary QCD axion and with fixed $m_a,~F_a,~\alpha$ and $\Omega_{\rm clump}/\Omega_{\rm DM}$ for axionlike particles.
The EROS-2 LMC survey observed no microlensing events, $N_{\rm obs}=0$, and thus, the parameter region $P(N_{\rm obs}=0,N_{\rm exp}) < 0.95$ corresponding to $N_{\rm exp} \gtrsim 3.0$ is excluded.

\begin{figure*}[t]
\begin{center}
\includegraphics[clip, width=7.5cm]{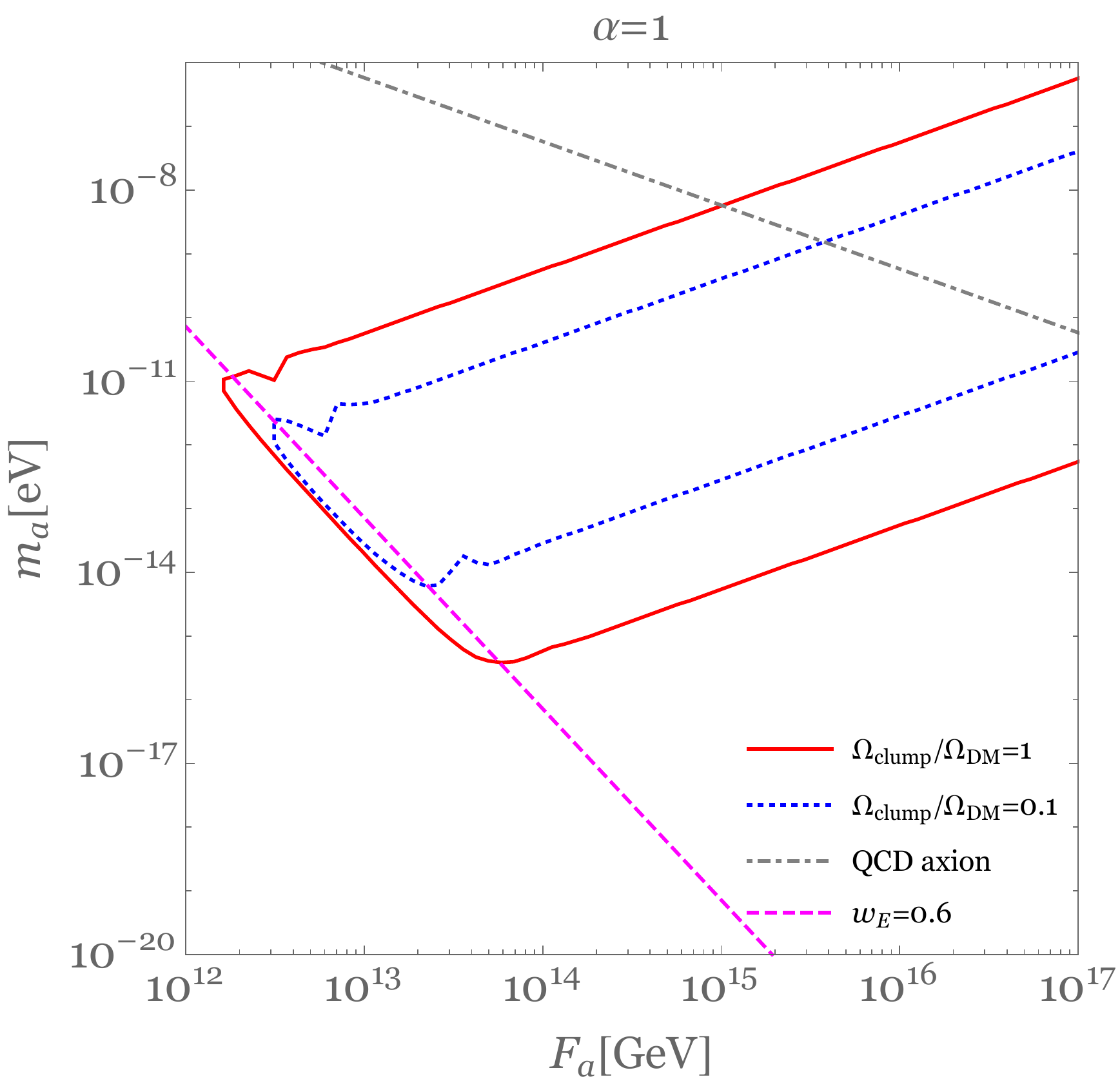}
\includegraphics[clip, width=7.5cm]{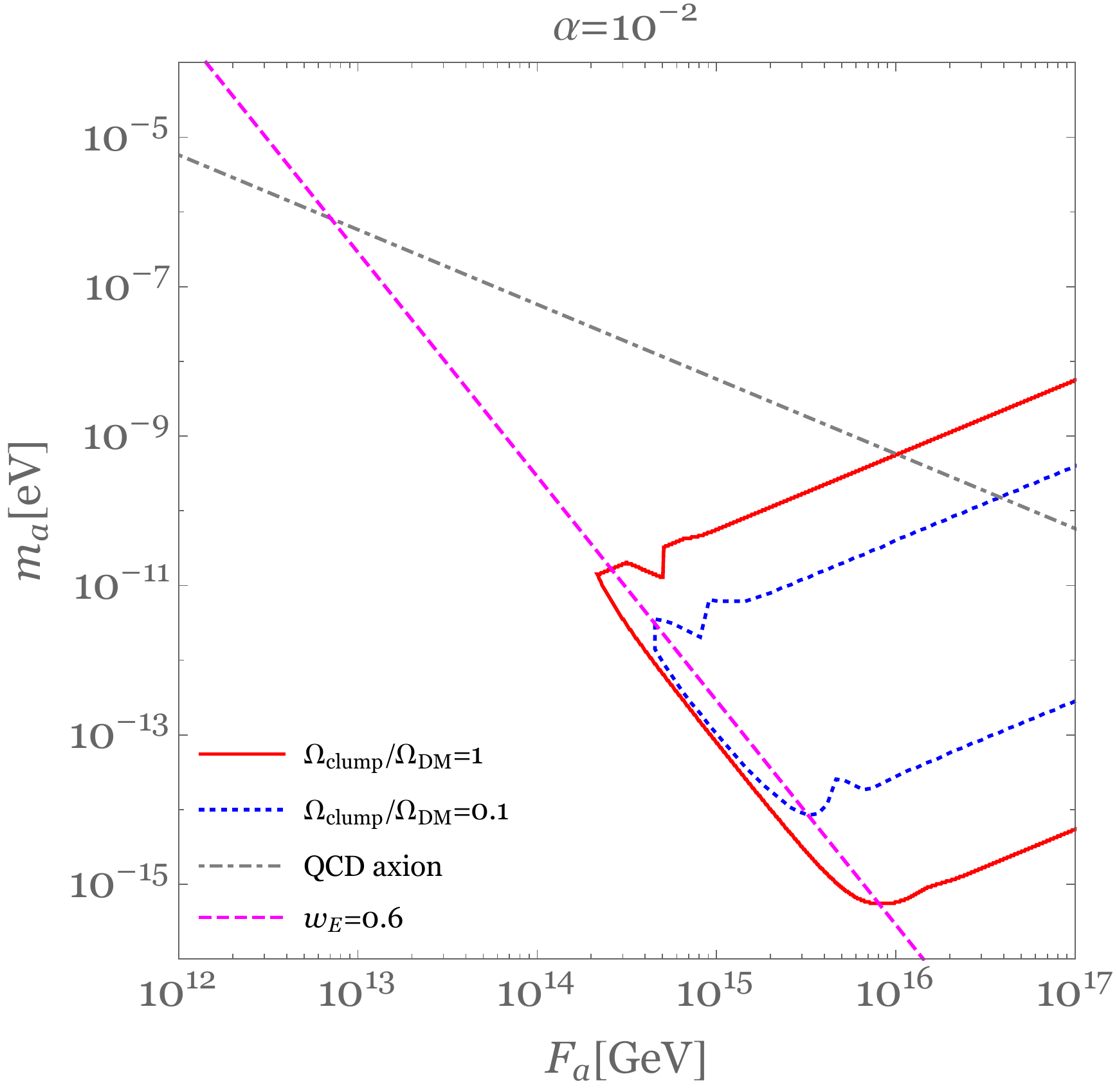}
\end{center}
\caption{Parameter regions on the $(F_a,m_a)$ plane excluded by the EROS-2 survey are shown for $\Omega_{\rm clump}/\Omega_{\rm DM}=1$ (red curve) and $\Omega_{\rm clump}/\Omega_{\rm DM}=0.1$ (blue dotted curve) with $\alpha =1 $ (left) and with $\alpha = 10^{-2}$ (right). The gray colored dotted-dashed contour and magenta colored dashed contour correspond to the parameter of the ordinary QCD axion and $w_E=0.6$, respectively. 
\label{fig:EROSmF}
}
\end{figure*}

Figure~\ref{fig:EROSmF} shows the parameter regions excluded by the EROS-2 survey on the $(F,m_a)$ plane for $\Omega_{\rm clump}/\Omega_{\rm DM}=1$ and $\Omega_{\rm clump}/\Omega_{\rm DM}=0.1$ with fixed $\alpha=1$.
The gray dotted-dashed line corresponds to the parameter region of the ordinary QCD axion.
The magenta colored dashed contour line corresponds to $w_E = 0.6$, where we use approximate expression of $w_E$ given by Eq.~\eqref{eq:wE}.
From the Fig.~\ref{fig:EROSmF}, we can find that the finite lens size effect becomes significant, and the constraint disappears around the contour $w_E\simeq 0.6$.
This is expected from Fig.~\ref{fig:impact parameter} since the threshold impact parameter becomes zero around $w_E\simeq 0.6$ as discussed in the previous section.
For a smaller fraction $\Omega_{\rm clump}/\Omega_{\rm DM}$, the width of the contour becomes narrower and eventually disappears.
When we focus on the QCD axion, a breaking scale much higher than the axion window can be constrained.

\begin{figure*}[!t]
\begin{center}
\includegraphics[clip, width=7.5cm]{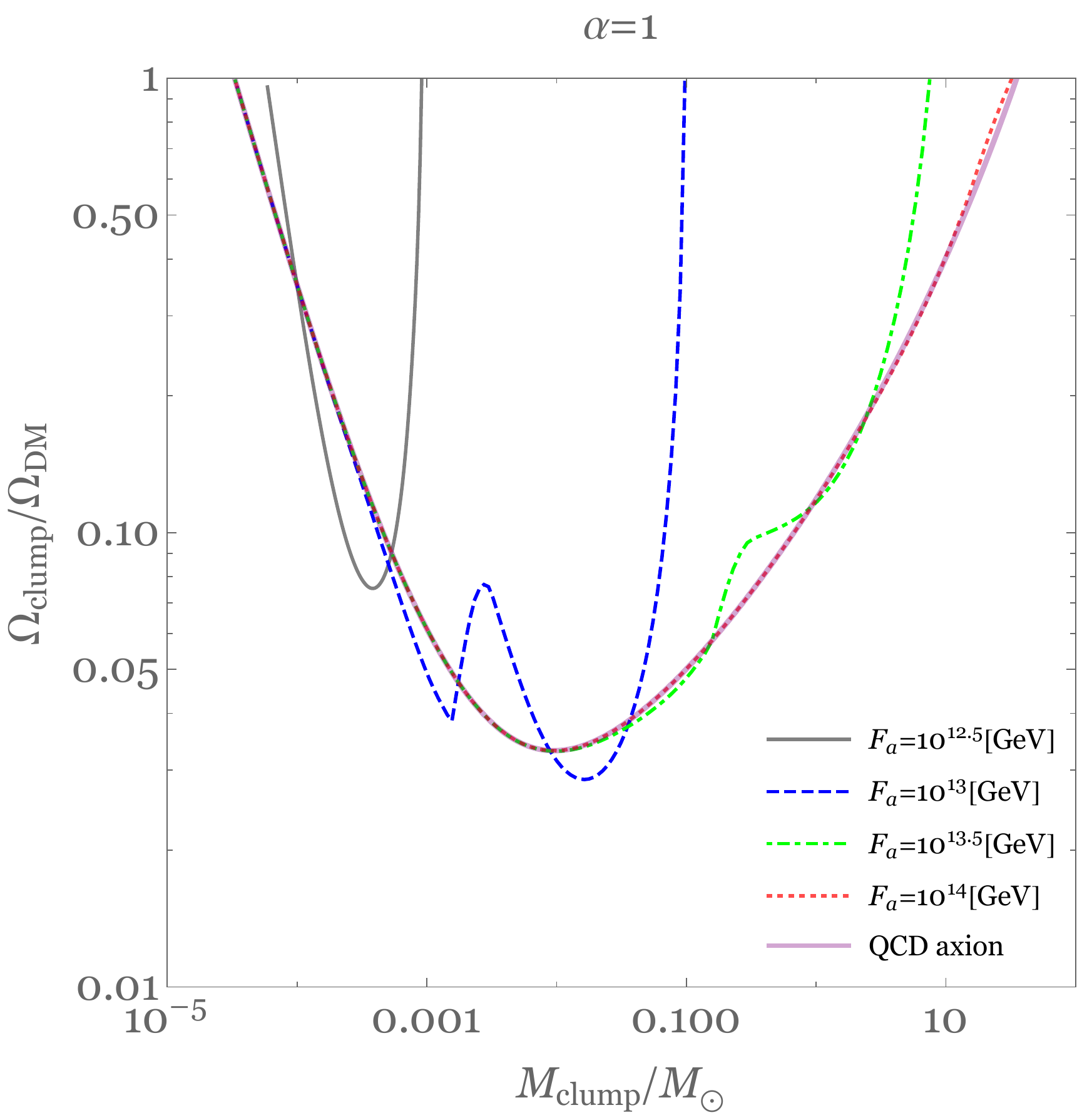}
\includegraphics[clip, width=7.5cm]{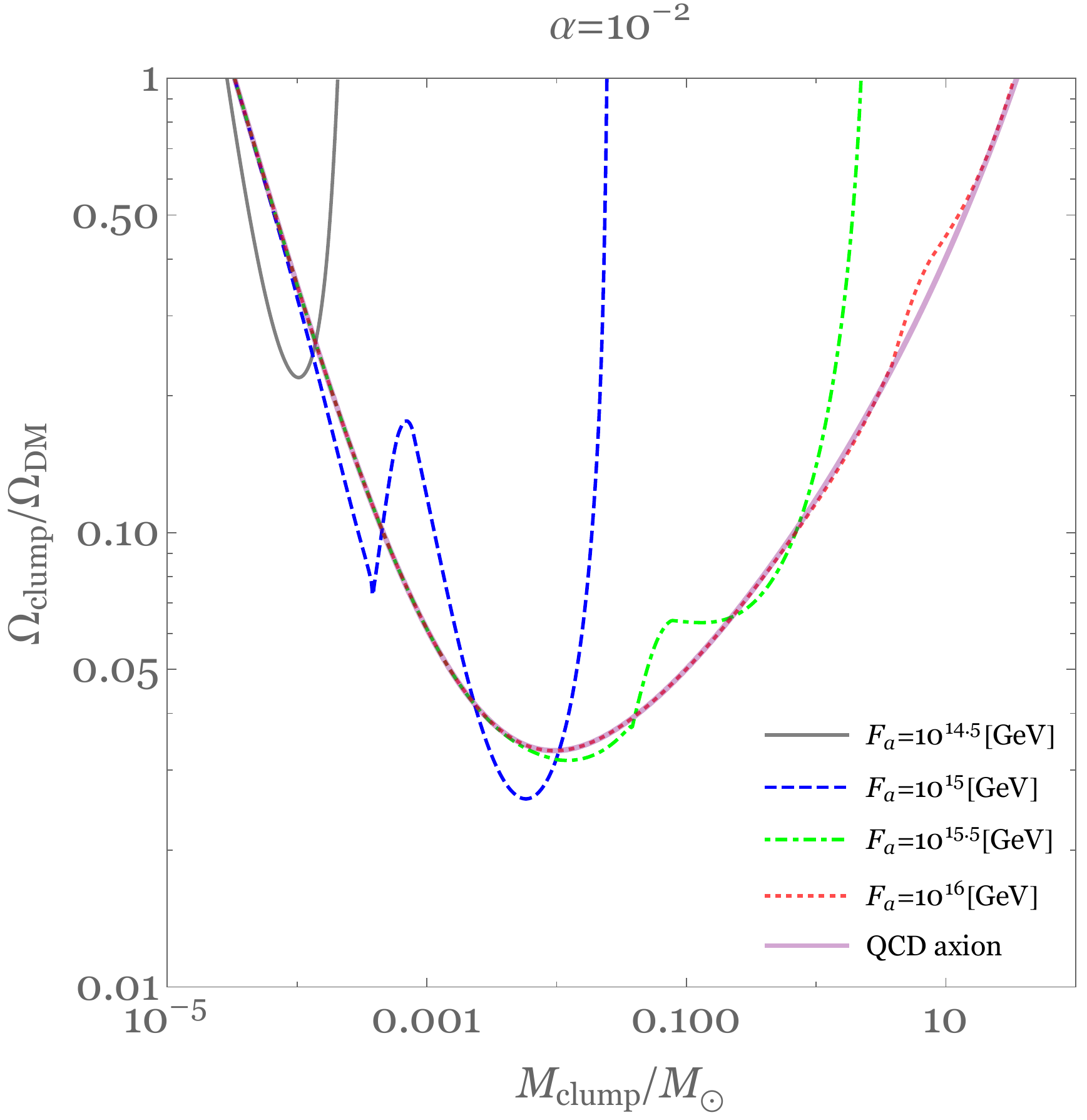}
\includegraphics[clip, width=7.5cm]{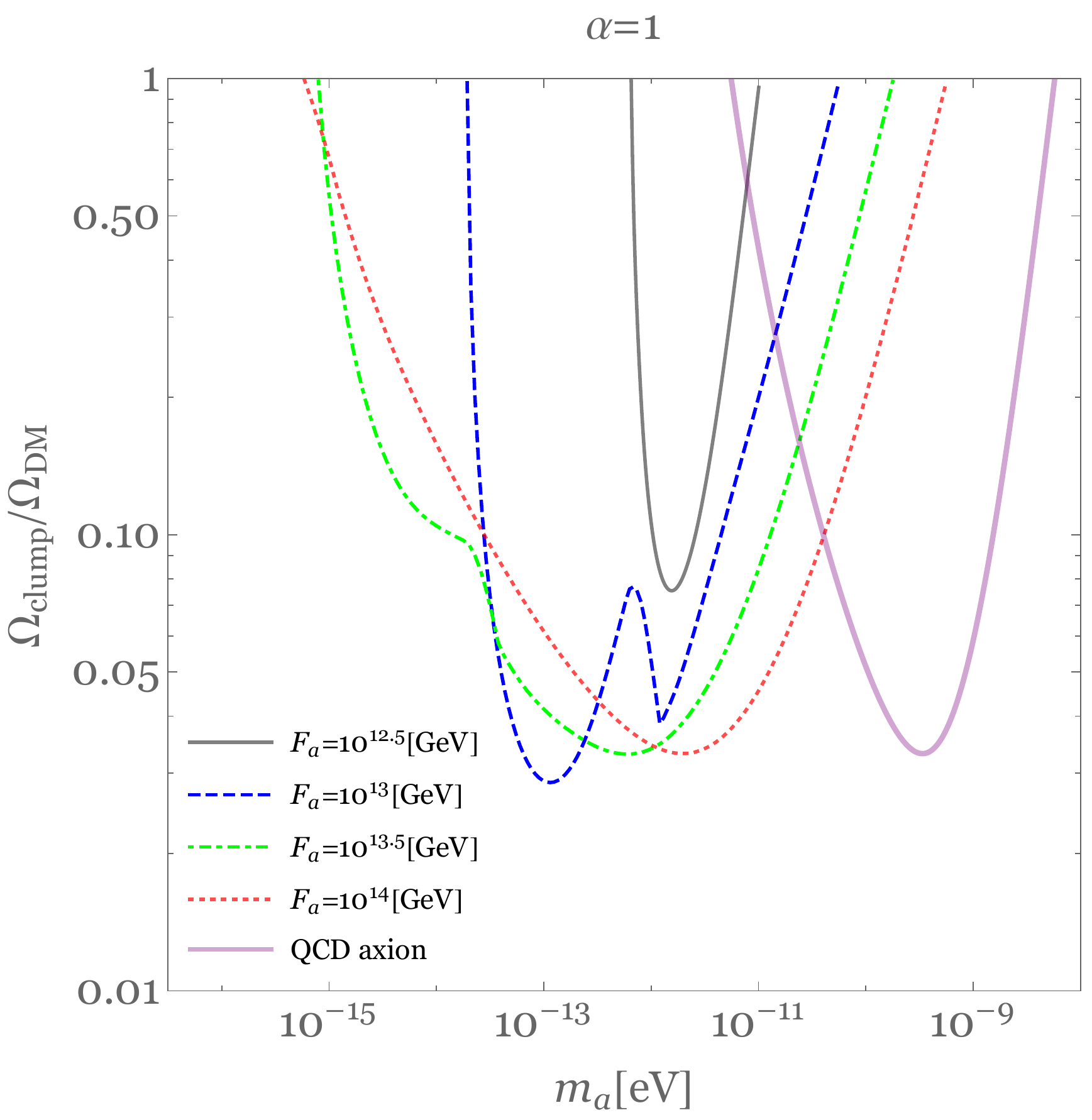}
\includegraphics[clip, width=7.5cm]{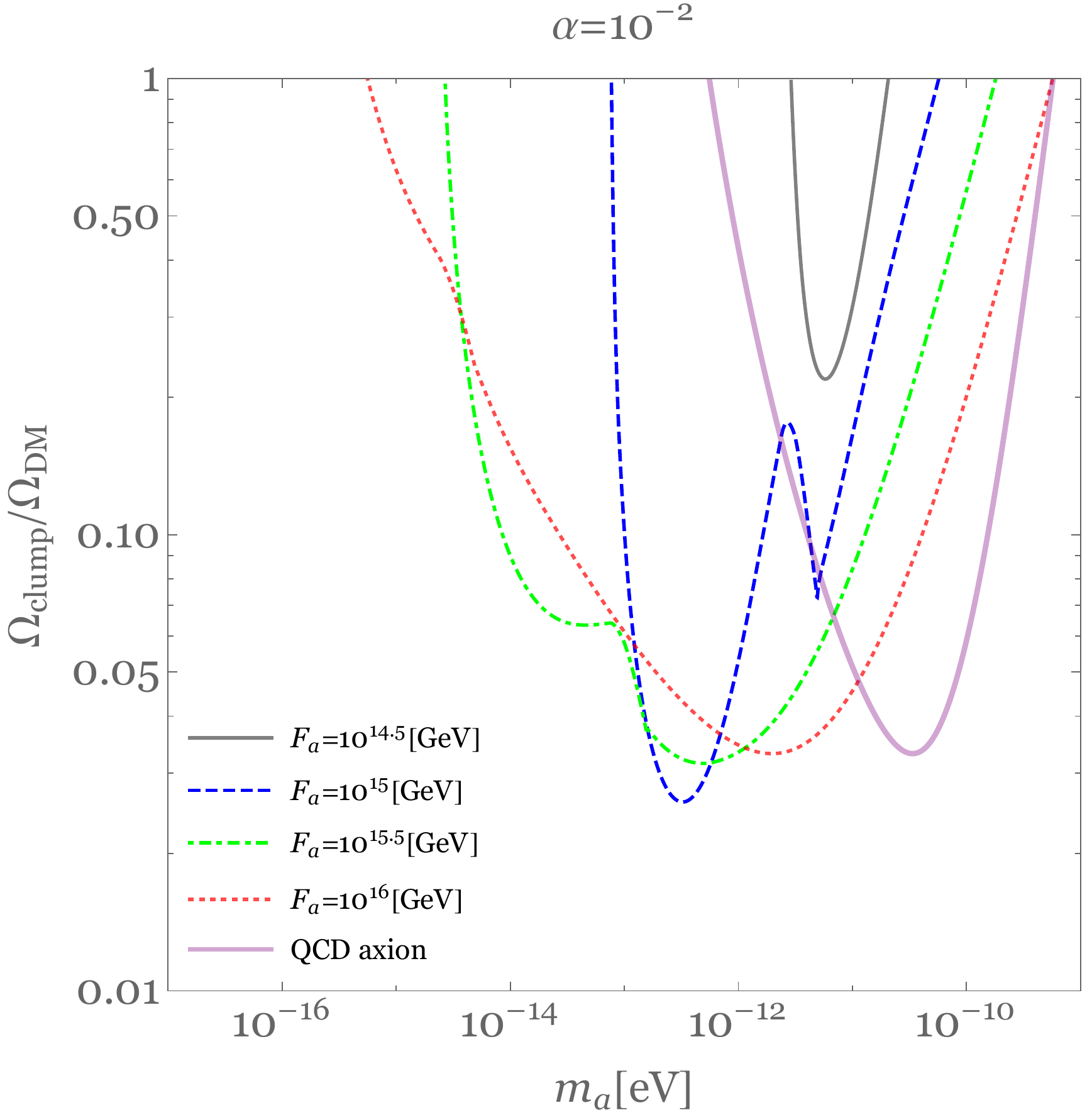}
\end{center}
\caption{Parameter regions on the $(m_a,\Omega_{\rm clump}/\Omega_{\rm DM})$ plane excluded by the EROS-2 survey are shown with $\alpha=1$ (upper left) and with $\alpha = 10^{-2}$ (upper right) for the axionlike particles with fixed $F_a$ and for the ordinary QCD axion. Excluded parameter regions on the $(m_a,\Omega_{\rm clump}/\Omega_{\rm DM})$ plane are shown for the same $F_a$ with $\alpha = 1$ (lower left) and with $\alpha  = 10^{-2}$ (lower right).
\label{fig:EROS DM fraction}
}
\end{figure*}

Figure~\ref{fig:EROS DM fraction} shows the parameter regions excluded by the EROS-2 survey on the $(M_{\rm clump}/M_{\odot},\Omega_{\rm clump}/\Omega_{\rm DM})$ plane (left) and on the $(m_a,\Omega_{\rm clump}/\Omega_{\rm DM})$ plane (right) with fixed $\alpha=1$ for several fixed breaking scales $F_a$ and for the ordinary QCD axion.
When the finite lens size effect is negligible, $10^{-4}\lesssim M_{\rm clump}/M_{\odot}\lesssim 10$ can be constrained by the EROS-2 survey, which is in agreement with the original results~\cite{Tisserand:2006zx,Green:2016xgy}.
When the finite lens size effect becomes important, contour curves have spikes as can be seen in the blue colored dashed and green colored dotted-dashed contour.
This behavior reflects that of the threshold impact parameter shown in Fig.~\ref{fig:impact parameter}.

Before closing this subsection, we comment on the finite source size effect.
We can approximately estimate $r_s$ by using $D_{\rm S}\sim D_{\rm L} \sim D_{\rm LS}$ with the assumption that source star radii are order of $R_{\odot}$.
As we confirmed, axion clumps whose masses are within $10^{-4}\lesssim M_{\rm clump}/M_{\odot}\lesssim 10$ are constrained in the EROS-2 survey.
In this parameter region, we find $r_s \ll 1$, and thus, we conclude that the finite source size effect is completely irrelevant.

\subsection{The Subaru HSC survey}\label{sec:Subaru HSC}

In this subsection, we give a microlensing constraint to the axion clump by using the observation data obtained by the Subaru Hyper Surprime-Cam (Subaru HSC).
A microlensing constraint on the PBH from the Subaru HSC observation is originally investigated in Ref.~\cite{Niikura:2017zjd}.
The microlensing constraint from the Subaru HSC observation on generic compact objects such as boson stars is investigated in Ref.~\cite{Croon:2020ouk} with including the finite source and lens size effects.

First of all, let us clarify the setup of the Subaru HSC survey.
The Subaru HSC survey focuses on source stars in MW and M31, whose distances from us are given by $D_S\simeq 770\,{\rm kpc}$.
Since M31 contains high dark matter density, microlensing events may occur not only inside MW but also inside the M31 itself.
The differential event rate is thus given by the sum of these: $d\Gamma=d\Gamma_{\rm MW}+d\Gamma_{\rm M31}$, where $d \Gamma_{\rm MW(M31)}$ is the differential event rate calculated in the MW (M31).
The circular velocity for the M31 is given in Ref.~\cite{Kafle:2018amm}, which is approximately given by $v_c\simeq 250\,{\rm km/s}$.
As done in Ref~\cite{Niikura:2017zjd}, spatial DM distributions in MW and M31 are assumed to be given by the NFW profile,
\begin{align}
\rho_{\rm NFW}= \dfrac{\rho_c}{(r/r_S)(1+r/r_S)^2},
\end{align}
where $r,r_S$ and $\rho_c$ are radii from the center, the scale radius, and the central density parameter of MW or M31, respectively.
For the MW, we have $r_S = 21.5\,{\rm kpc},~\rho_c=0.184\,{\rm GeV/cm^{3}}$, and $r=r_{\rm MW}$ given by 
\begin{align}
r_{\rm MW}(D_L)= \sqrt{R_{\rm sun}^2 -2R_{\rm sun} D_L\cos{l'}\cos{b'}+D_L^2 },
\end{align}
where $(l',b')=(\ang{121.2},\ang{-21.6})$~\cite{Klypin:2001xu}.
For M31, we have $r_S=25\,{\rm kpc},~\rho_c=0.19\,{\rm GeV/cm^{3}}$, and $r=r_{\rm M31}$ is given by
\begin{align}
r_{\rm M31} (D_L)= D_S -D_L.
\end{align}
The number of stars used in the Subaru HSC survey, $N_{\rm star}=8.7\times 10^7$, with the observation time $T_{\rm obs}=7$ hours, gives the exposure time $E=N_{\rm star}T\simeq 7.0\times 10^4$  sidereal years.
The detection efficiency is given by Fig.~19 in Ref.~\cite{Niikura:2017zjd} in terms of full-width-half-maximum (FWHM) timescale, $t_{\rm FWHM}$.
In our analysis, for simplicity, we use $t_{\rm FWHM}=\hat{t}$ and approximate the detection efficiency as $\epsilon =0.5$ in the region with
$2\,{\rm min}\leq \hat{t} \leq 7\,{\rm hr}$.

As we see later, the constrained mass of an axion clump in the Subaru HSC survey is much lighter than that in the EROS-2 survey.
Hence, the Einstein ring radius is smaller, and thus, the finite source size corrections are important, different from the EROS-2 survey.
Therefore, we must include the finite source size effect otherwise the microlensing constraint is overestimated.
To estimate $N_{\rm exp}$, we need a distribution of source star radii.
In our analysis, we use the distribution of source star radii shown by Fig. 4 in Ref.~\cite{Smyth:2019whb}.
With this setup, the Subaru HSC observation observed single microlensing event candidate, that is $N_{\rm obs}=1$, and hence, the parameter region $P(N_{\rm obs}=0,N_{\rm exp})+P(N_{\rm obs}=1,N_{\rm exp})<0.05$ corresponding to $N_{\rm exp}\gtrsim 4.74$ is excluded with a $95\%$ confidence level.

\begin{figure*}[!t]
\begin{center}
\includegraphics[clip, width=7.5cm]{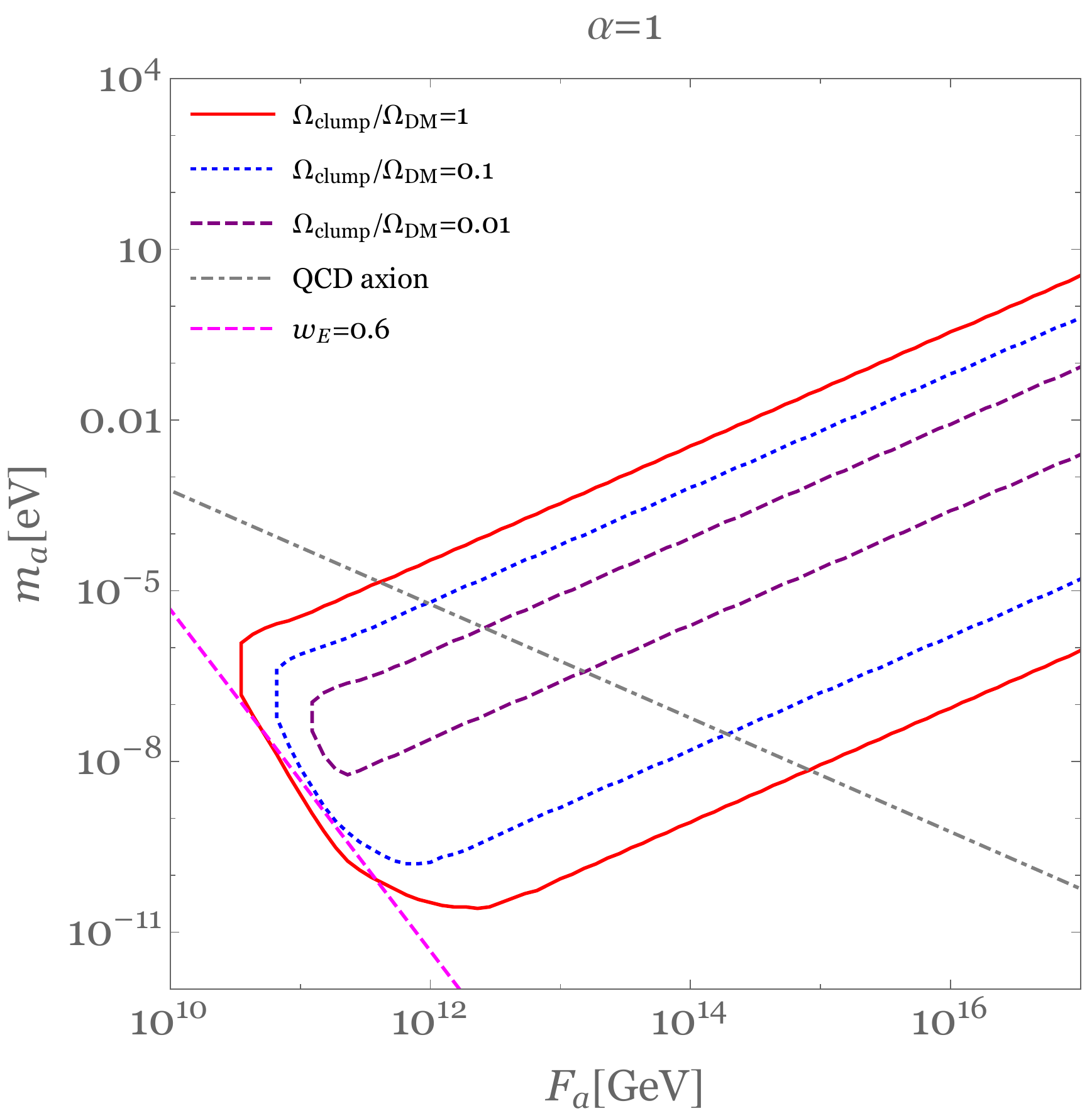}
\includegraphics[clip, width=7.5cm]{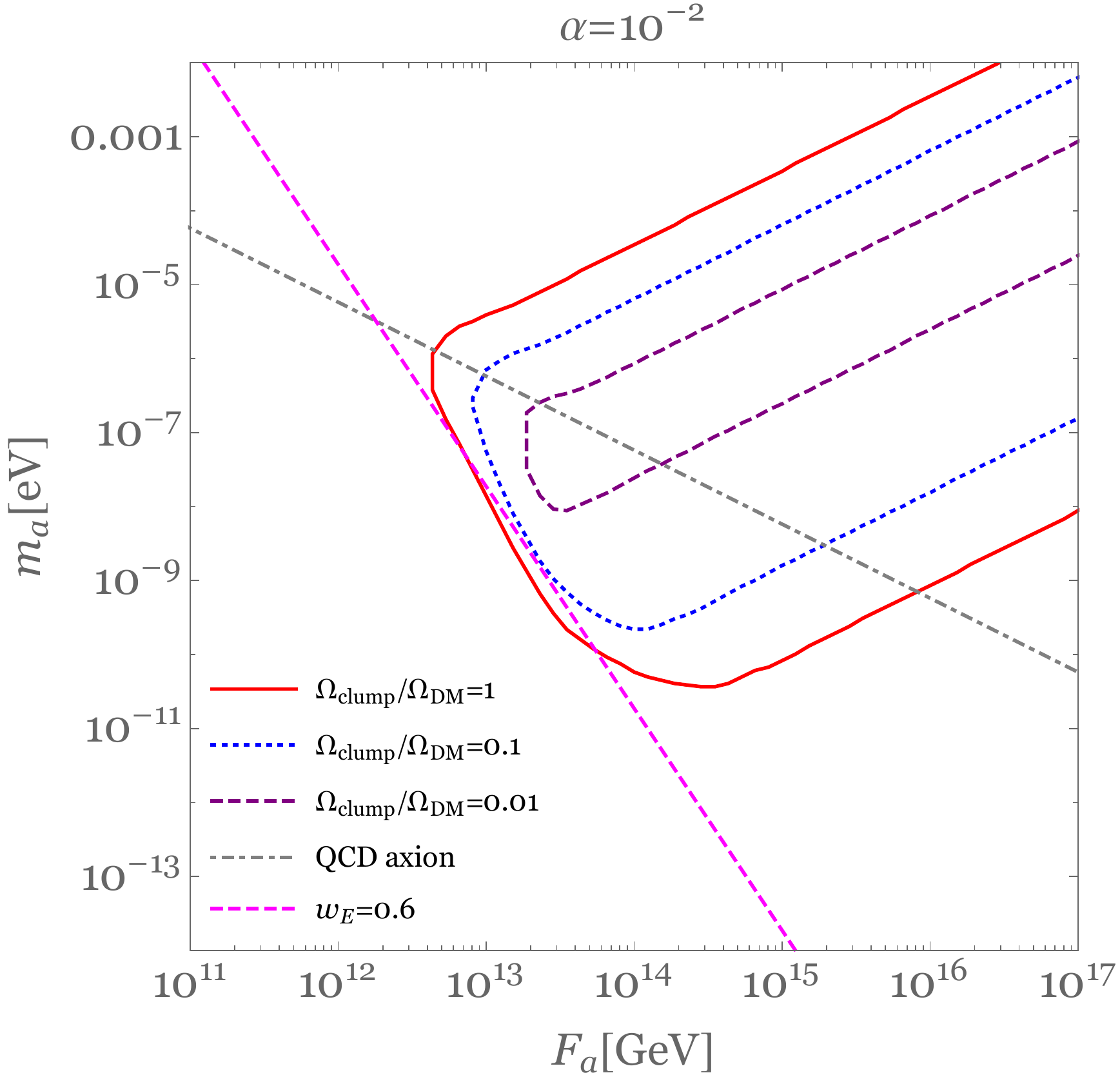}
\end{center}
\caption{Parameter regions on the $(F_a,m_a)$ plane excluded by the Subaru HSC observation are shown for $\Omega_{\rm clump}/\Omega_{\rm DM}=1$ (red curve), $\Omega_{\rm clump}/\Omega_{\rm DM}=0.1$ (blue dotted curve), and $\Omega_{\rm clump}/\Omega_{\rm DM}=0.01$ (purple dashed curve) with $\alpha =1 $ (left) and with $\alpha = 10^{-2}$ (right). The gray colored dotted-dashed contour and magenta colored dashed contour correspond to the parameter of the ordinary QCD axion and $w_E=0.6$, respectively. 
\label{fig:SubarumF}
}
\end{figure*}

Figure~\ref{fig:SubarumF} shows the parameter region excluded by the Subaru HSC observation on the $(F_a,m_a)$ plane for $\Omega_{\rm clump}/\Omega_{\rm DM}=1,~\Omega_{\rm clump}/\Omega_{\rm DM}=0.1$, and $\Omega_{\rm clump}/\Omega_{\rm DM}=0.01$ with fixed $\alpha=1$.
The magenta colored dashed contour line corresponds to $w_E = 0.6$, where we use the approximate expression of $w_E$ given by Eq.~\eqref{eq:wE}.
Similar to the EROS-2 survey discussed in the previous subsection, the finite lens size effect becomes important, and the constraint disappears around $w_E\simeq 0.6$.
Also, a smaller fraction $\Omega_{\rm clump}/\Omega_{\rm DM}$ makes the width of the contour narrower.
In comparison to the EROS-2 survey, a smaller breaking scale of the QCD axion can be constrained.
In particular, for $\alpha =1$, axion clumps composed of the QCD axion can be constrained around $F_a\gtrsim 10^{12}\,{\rm GeV}$.

\begin{figure*}[t]
\begin{center}
\includegraphics[clip, width=7.5cm]{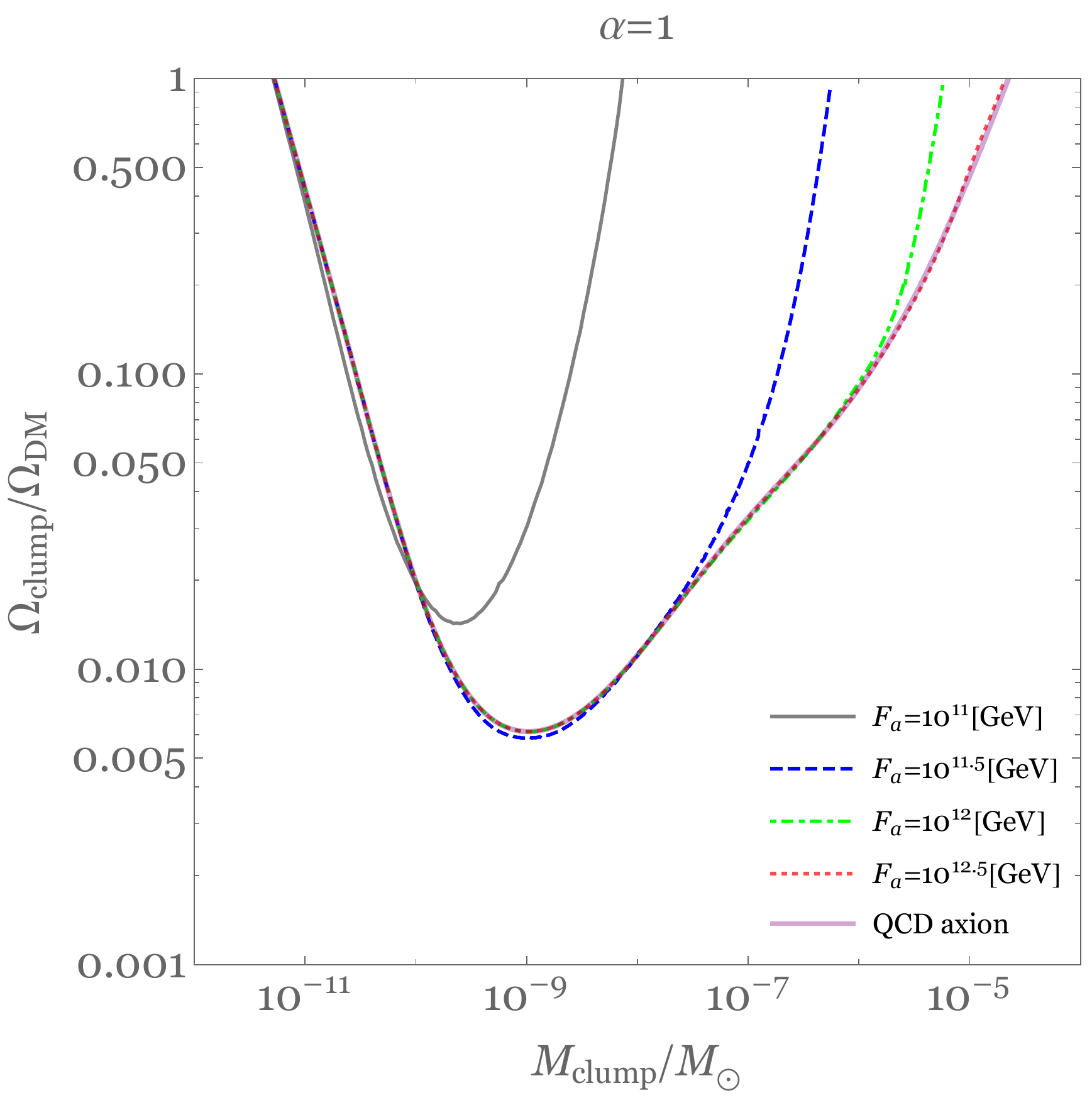}
\includegraphics[clip, width=7.5cm]{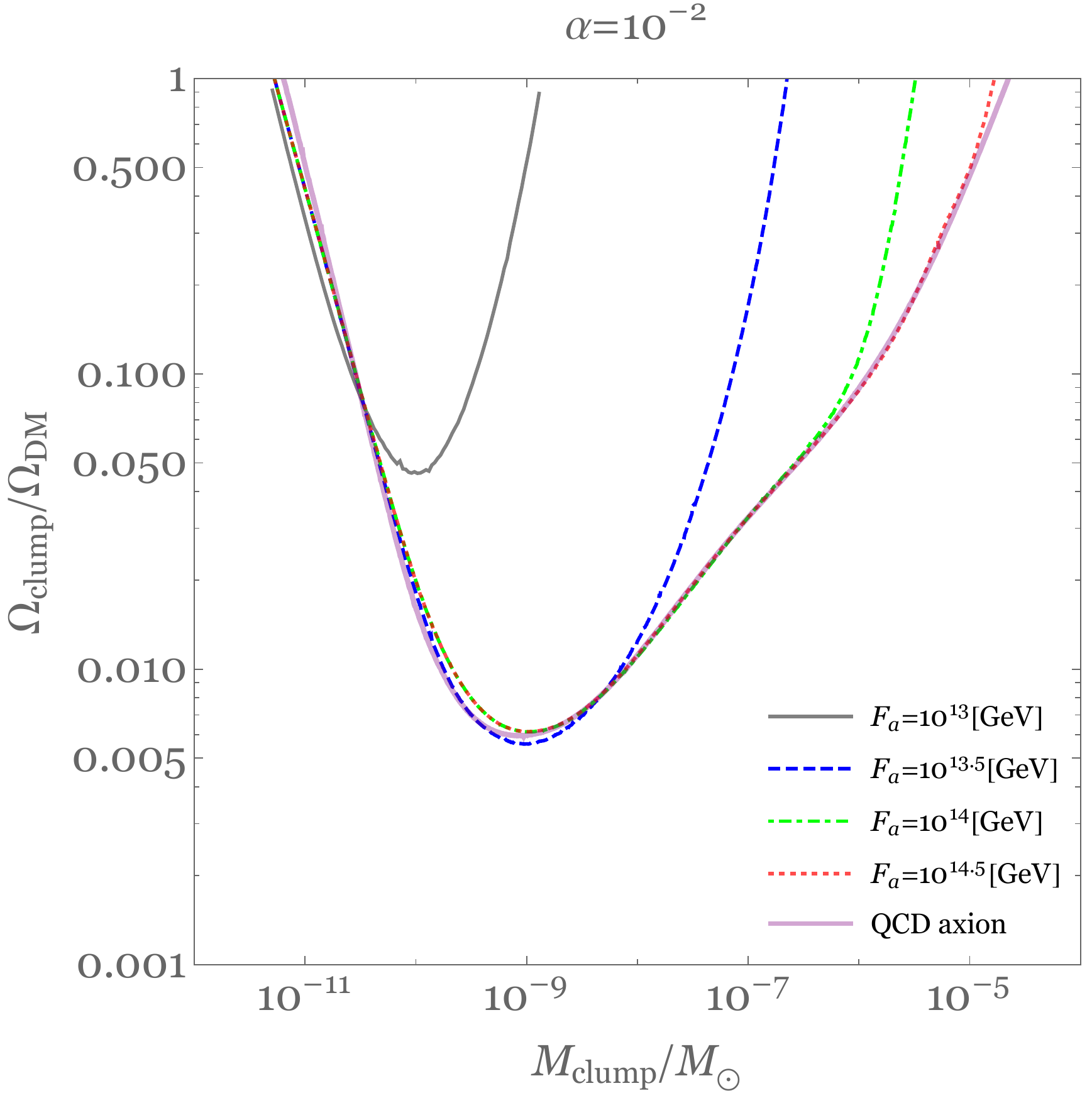}
\includegraphics[clip, width=7.5cm]{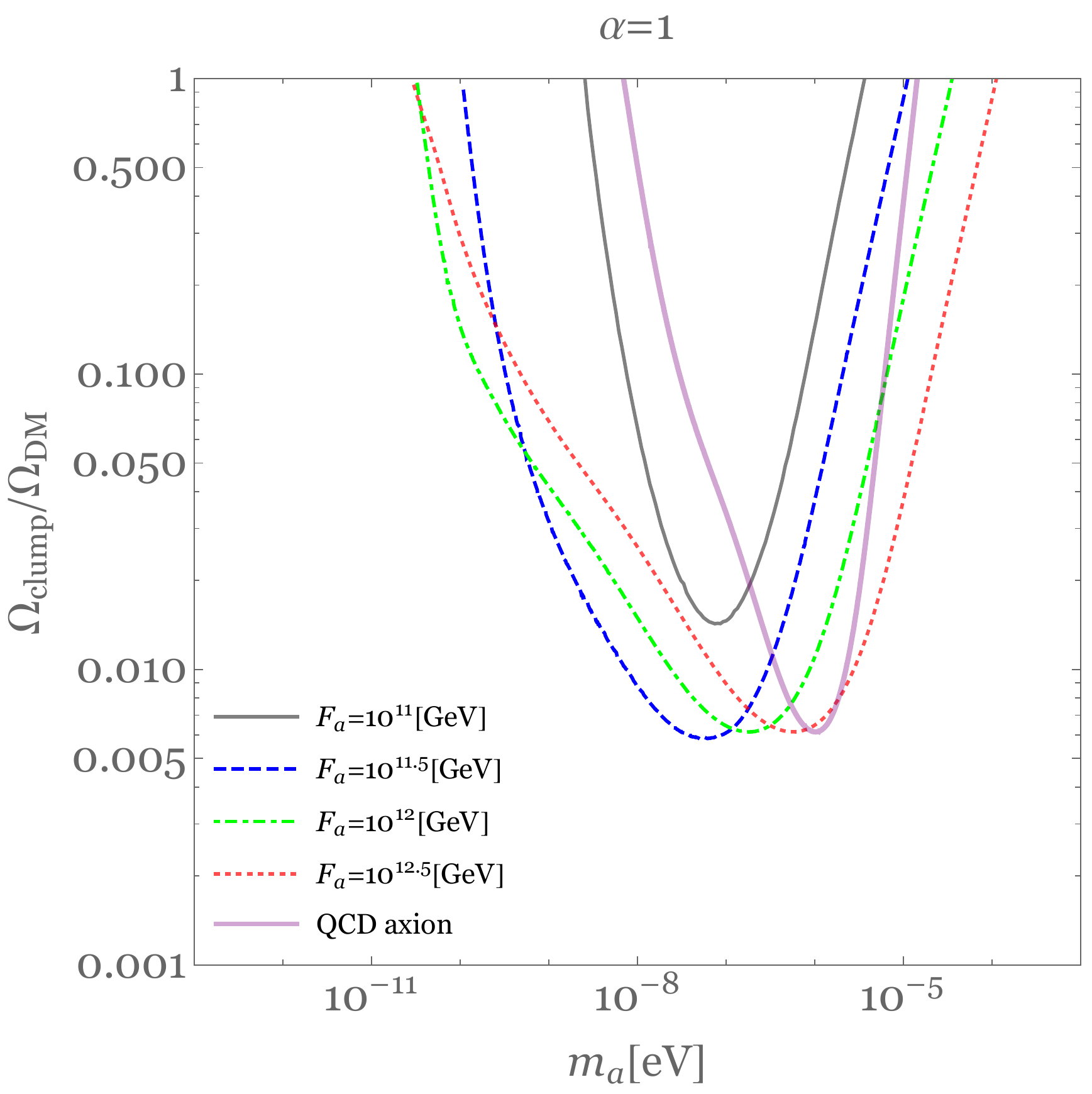}
\includegraphics[clip, width=7.5cm]{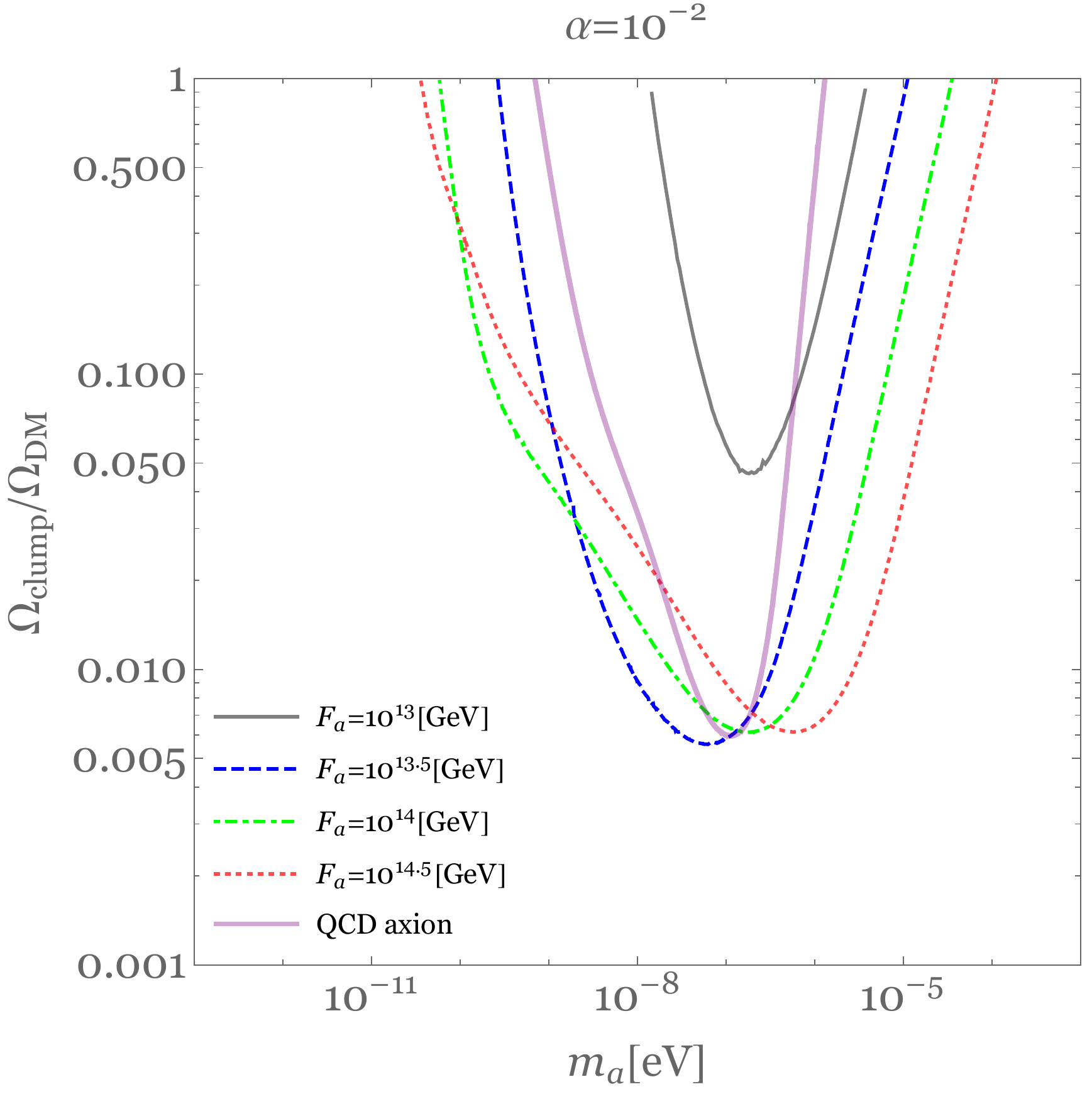}
\end{center}
\caption{Parameter regions on the $(m_a,\Omega_{\rm clump}/\Omega_{\rm DM})$ plane excluded by the Subaru HSC observation are shown with $\alpha=1$ (upper left) and with $\alpha = 10^{-2}$ (upper right) for the axionlike particles with several fixed $F_a$ and the ordinary QCD axion. Excluded parameter regions on the $(m_a,\Omega_{\rm clump}/\Omega_{\rm DM})$ plane are shown for the same $F_a$ with $\alpha = 1$ (lower left) and with $\alpha  = 10^{-2}$ (lower right).
\label{fig:Subaru DM fraction}
}
\end{figure*}

Figure~\ref{fig:Subaru DM fraction} shows the parameter region excluded by the Subaru HSC observation on the $(M_{\rm clump}/M_{\odot},\Omega_{\rm clump}/\Omega_{\rm DM})$ plane (left) and on the $(m_a,\Omega_{\rm clump}/\Omega_{\rm DM})$ plane (right) with fixed $\alpha=1$ for several fixed breaking scales $F_a$ and for the ordinary QCD axion.
When the finite lens size effect is negligible, $10^{-11}\lesssim M_{\rm clump}/M_{\odot}\lesssim 10^{-5} $ can be constrained.
It should be noted that a lower bound of $M_{\rm clump}$ is cut off by the finite source size effect where the pointlike Einstein ring radius is comparable to the source star radius in that region.

\section{Repulsive self-interactions}\label{repulsive}

In the previous sections, we discussed the microlensing constraints on clumps composed of axions where the quartic coupling is negative, $\lambda <0$, leading to the attractive self-interaction.
In this section, we consider the microlensing constraints on clumps composed of generic light scalar fields whose potentials are described by
\begin{align}
V(\phi) = \frac{1}{2}m_r^2 \phi^2 +\frac{\lambda_r}{4!}\phi^4,
\end{align}
with $\lambda_r>0$ leading to repulsive self-interactions.

An analysis of a spherically symmetric classical field configuration in the nonrelativistic regime is essentially the same as the case with the attractive self-interaction.
It is convenient to parametrize the tiny positive quartic coupling $\lambda_r$ in term of $F_r$ defined by the relation $\lambda_r \equiv m_r^4/F_r^4$.
We also assume a linear $\times$ exponential ansatz for the clump configuration,
\begin{align}
\Phi(r) = \sqrt{\frac{N}{7\pi R_r^3}}\left(1+\frac{r}{R_r}\right) e^{-\frac{r}{R_r}},\label{eq:ansatz repulsive}
\end{align}
where $R_r$ is the length scale, which controls the shape of the clump.
Then, the Hamiltonian of the clump is given by
\begin{align}
\widetilde{H} \simeq a\frac{\widetilde{N}}{\widetilde{R}_r^2}-b\frac{\widetilde{N}^2}{\widetilde{R}_r}+c\frac{\widetilde{N}^2}{\widetilde{R}_r^3},
\end{align}
where constants $a,~b$, and $c$ are given by Eq.~\eqref{eq:constants}, while $\widetilde{N}$ and $\widetilde{R_r}$ are defined by 
\begin{align}
	&\widetilde{N} \equiv \frac{m_r^2 \sqrt{G_N}}{F_r}N,\\
	&\widetilde{R}_r \equiv \frac{m_r}{F_r^3\sqrt{G_N}}\widetilde{R_{r}}.
\end{align}
From the above Hamiltonian, we obtain the extremum of $\widetilde{R}$ given by
\begin{align}
\widetilde{R}_r =\frac{a+\sqrt{a^2+3bc \widetilde{N}^2}}{b\widetilde{N}}.
\end{align}
The other branch is an unphysical solution, which has a negative radius.
In the presence of the repulsive self-interactions, the clump can have an arbitrary large particle number $\widetilde{N}$.
Similar to the attractive self-interaction case, it is convenient to express $\widetilde{R}_r$ in term of $\alpha$ and $\widetilde{R}_{\rm min}$ parameters defined by Eq.~\eqref{eq:alpha definition} as
\begin{align}
&\widetilde{N}\equiv \alpha \widetilde{N}_r,~\widetilde{N}_r\equiv \frac{a}{\sqrt{3bc}},\nonumber \\
&\widetilde{R_r} = \frac{1}{\alpha}(1+\sqrt{1+\alpha^2}) \widetilde{R}_{\rm min}.
\end{align}

In comparison to the case with the attractive self-interaction, $\alpha$ can become arbitrarily large and the minimum size of the clump, $\widetilde{R}_{\rm min}$, is realized for $\alpha \to \infty$.
For fixed particle number $\widetilde{N}$, clump configurations are slightly different between the positive and negative quartic couplings (see Fig.~\ref{fig:branches} in Sec.~\ref{sols}).
Note that the clump configuration with a repulsive self-interaction is very similar to that with the attractive self-interaction for $\alpha \leq 1$.
Hence, microlensing constraints on clumps with a repulsive self-interaction are almost the same as that with the attractive self-interaction in that regime.
For this reason, we investigate microlensing constraints on clumps with a  repulsive self-interaction for large $\alpha\gg 1$.

\subsection{Microlensing constraints on clumps with repulsive self-interactions}

In this subsection, we give microlensing constraints on clumps with repulsive self-interactions.
Since the clump configuration is approximated by a linear $\times$ exponential ansatz, the threshold impact parameter including the finite lens and source size effects is the same as that in the attractive self-interaction case, which is shown in Fig.~\ref{fig:finite_size_source}.
By computing the expected number of events, $N_{\rm exp}$, defined by Eq.~\eqref{eq:expected number of events} with the setups of the EROS-2 survey and the Subaru HSC observation explained in Secs.~\ref{sec:EROS-2} and \ref{sec:Subaru HSC}, we can give microlensing constraints.

\begin{figure*}[t]
\begin{center}
\includegraphics[clip, width=7.5cm]{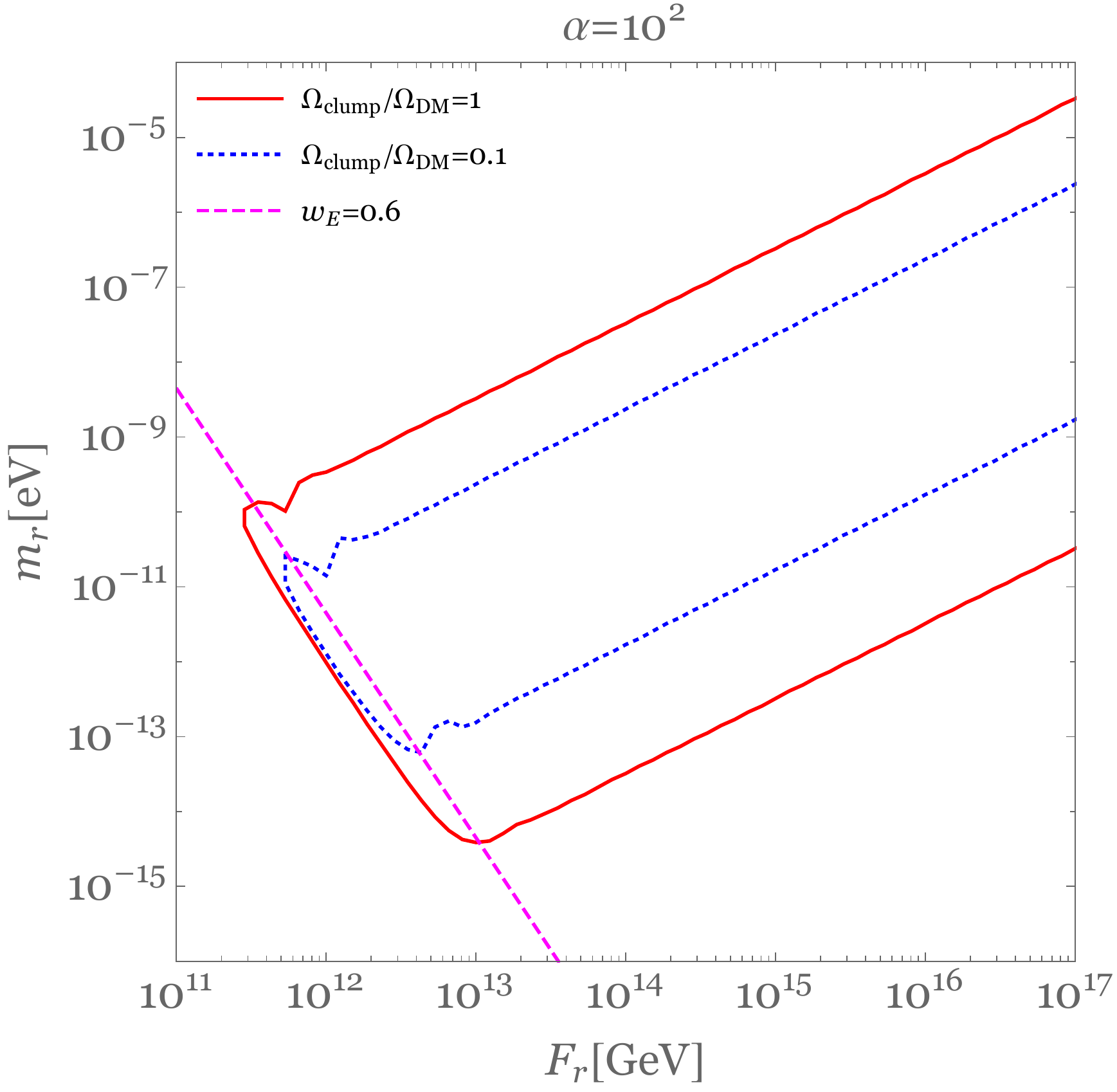}
\includegraphics[clip, width=7.5cm]{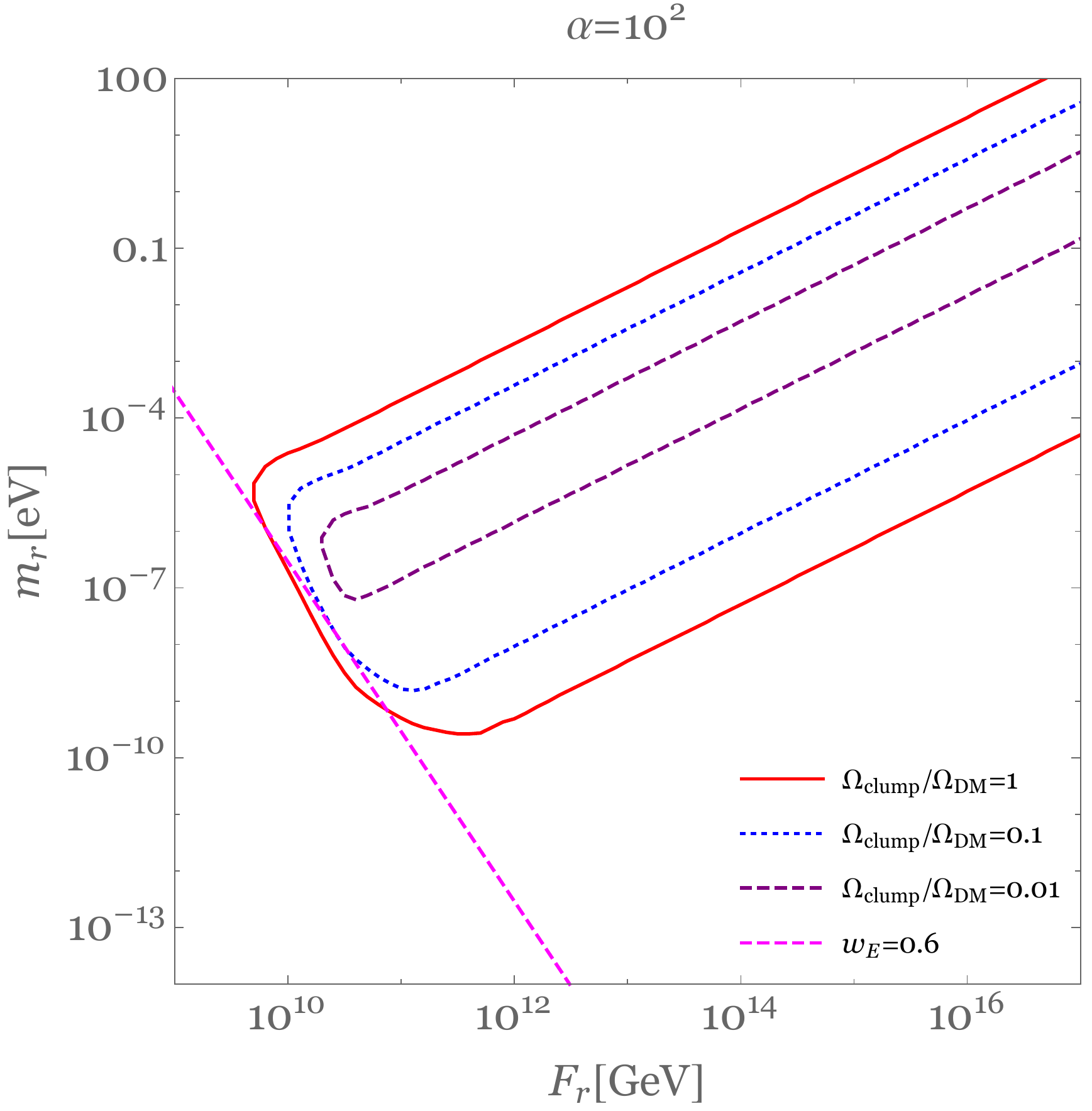}
\end{center}
\caption{Parameter regions on the $(F_r,m_r)$ plane excluded by the EROS-2 survey (left) and (right) are shown for $\Omega_{\rm clump}/\Omega_{\rm DM}=1$ (the red curve), $\Omega_{\rm clump}/\Omega_{\rm DM}=0.1$ (the blue dotted curve), and $\Omega_{\rm clump} / \Omega_{\rm DM}=0.01$ (the purple dashed curve) with $\alpha =10^{2} $. The magenta colored dashed contour correspond to $w_E=0.6$.
\label{fig:repulsive_contour}
}
\end{figure*}

In Fig.~\ref{fig:repulsive_contour}, we show microlensing constraints from the EROS-2 survery and the Subaru HSC observations on the scalar clump with a positive quartic coupling in the $(m_r,F_r)$ plane for $\Omega_{\rm clump}/\Omega_{\rm DM}=1,~0.1,~0.01$ with fixed $\alpha = 10^{2}$.
We also draw a contour line of $w_E = 0.6$ with assuming $D_{\rm S}\sim D_{\rm LS}\sim D_{\rm L}$ for a repulsive self-interaction.
As is the same as the attractive self-interaction case, the finite lens size effect gives a significant effect, and microlensing constraints disappear at around $w_E=0.6$.
Since we take $\alpha = 10^2$, a heavier mass of the scalar field, $m_r$, and a higher breaking scale, $F_r$, are more constrained compared to the results for the attractive self-interaction with $\alpha = 1,~10^{-2}$.
\begin{figure*}[!t]
\begin{center}
\includegraphics[clip, width=7.5cm]{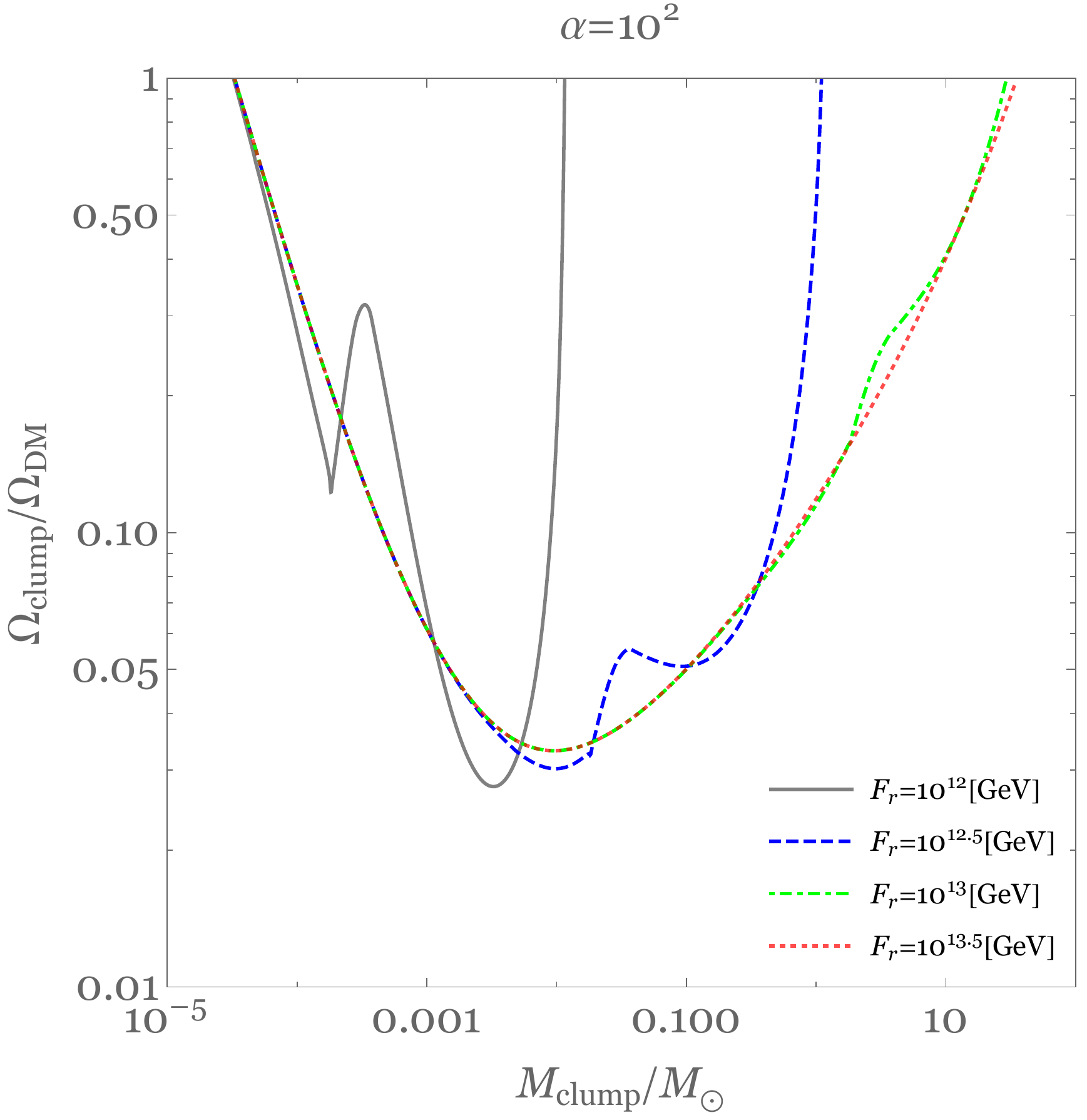}
\includegraphics[clip, width=7.5cm]{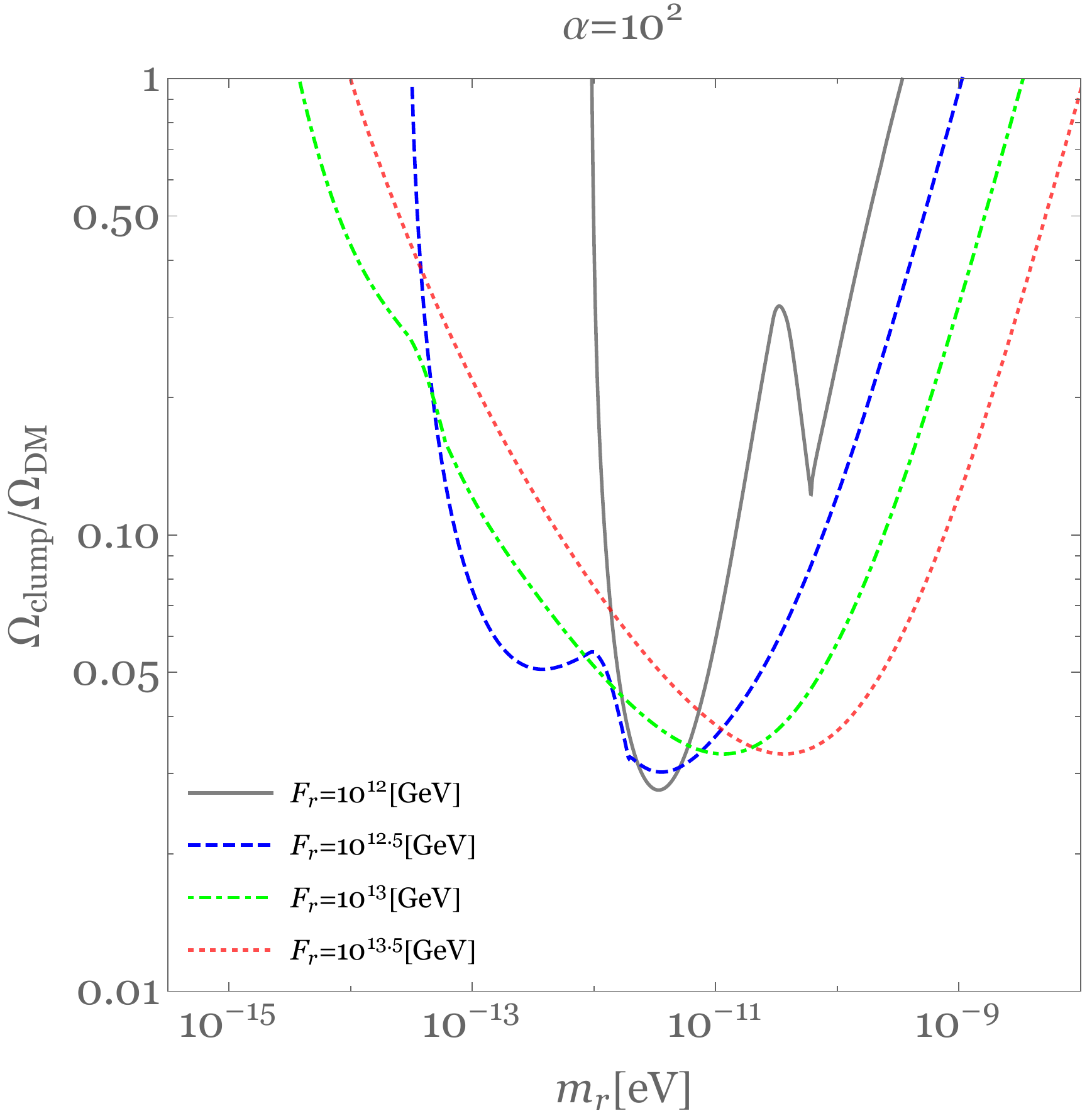}
\includegraphics[clip, width=7.5cm]{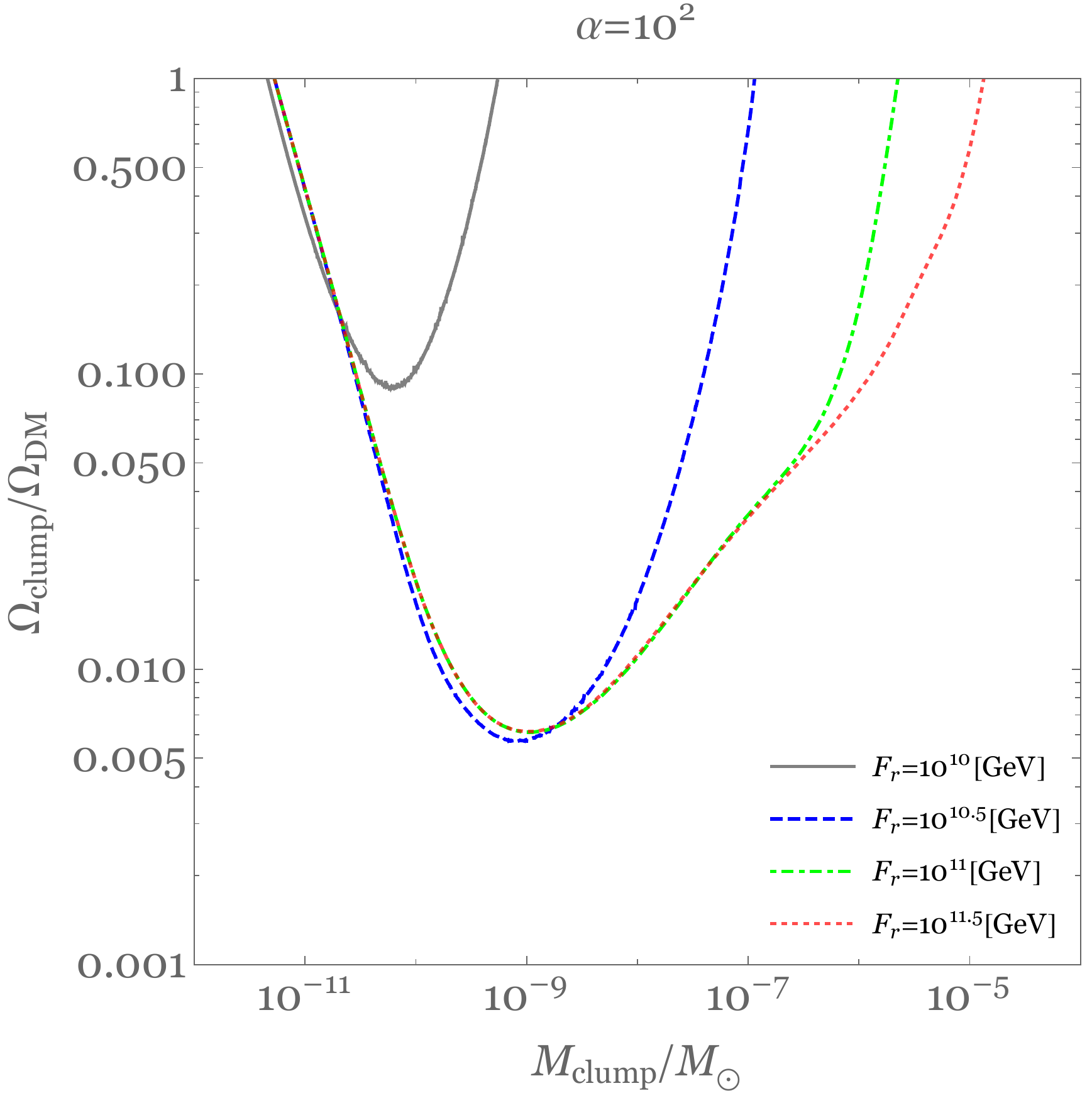}
\includegraphics[clip, width=7.5cm]{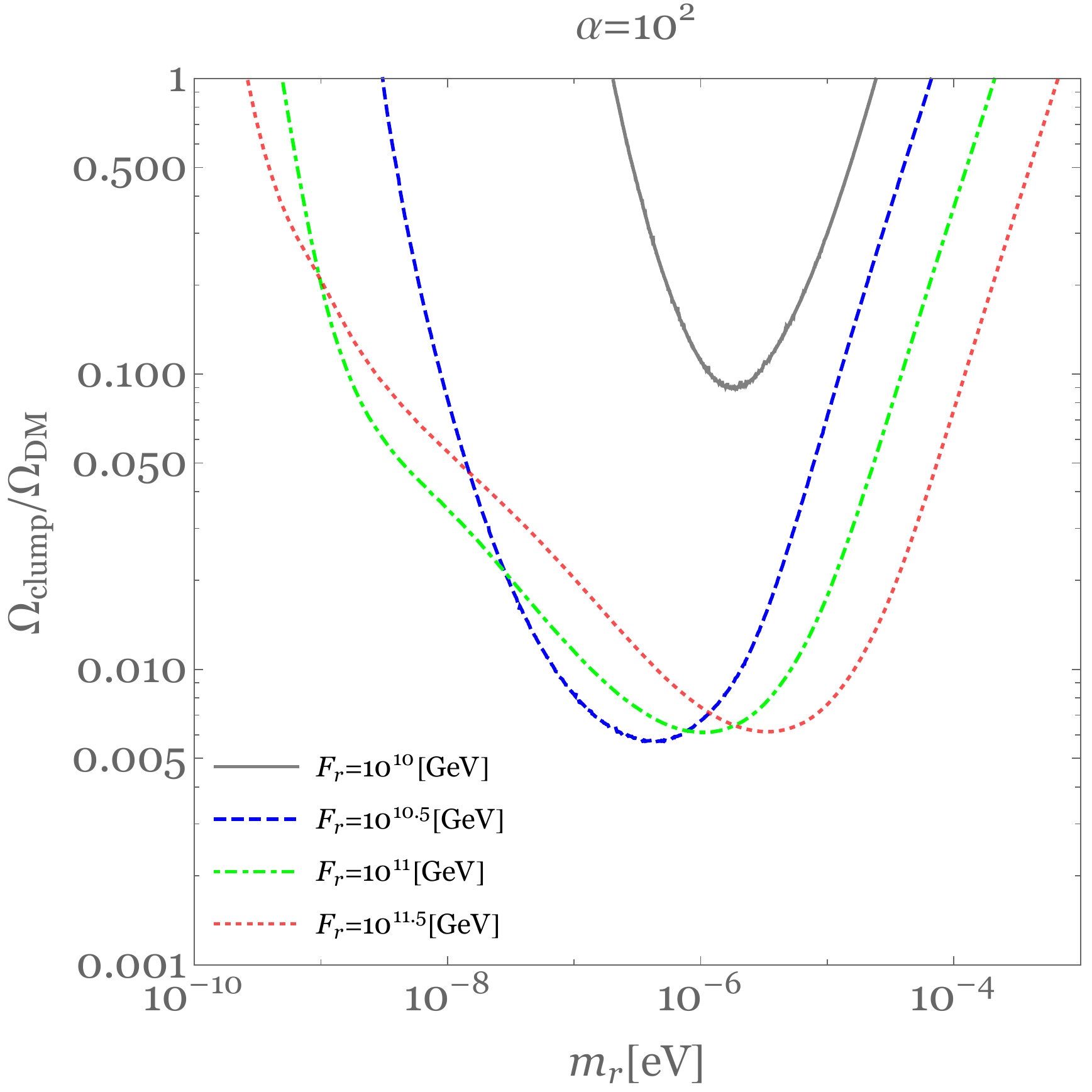}
\end{center}
\caption{Parameter regions excluded by the EROS-2 survey on the $(M_{\rm clump}/M_{\odot},\Omega_{\rm clump}/\Omega_{\rm DM})$ plane (upper left) and on the $(m_r, \Omega_{\rm clump}/\Omega_{\rm DM})$ plane (upper right) are shown with $\alpha = 10^2$ for several fixed $F_r$.
Parameter regions excluded by the Subaru HSC observations on the $(M_{\rm clump}/M_{\odot}, \Omega_{\rm clump}/\Omega_{\rm DM})$ plane (lower left) and on the $(m_r,\Omega_{\rm clump}/\Omega_{\rm DM})$ plane (lower right) are shown for $\alpha = 10^2$ for fixed $F_r$.
\label{fig:repulsive DM fraction}
}
\end{figure*}
In Fig.~\ref{fig:repulsive DM fraction}, we show microlensing constraints on the $(M_{\rm clump}/M_{\odot},\Omega_{\rm clump}/\Omega_{\rm DM})$ plane and on the $(m_r,\Omega_{\rm clump},\Omega_{\rm DM})$ plane for several fixed $F_r$ with fixed $\alpha =10^2$.
As is the same as the attractive self-interaction case, the finite source size effect is not important in the EROS-2 survey, while it becomes important in the Subaru HSC observation.

\section{Conclusions}\label{sec:conclusion}

In this paper, we have derived microlensing constraints from the EROS-2 survey and the Subaru HSC observations on spherically symmetric axion clumps (``boson stars") composed of the ordinary QCD axion and axionlike particles.
Since axion clumps are not generally, pointlike massive objects, (to be contrasted with PBHs), we calculated the threshold impact parameter by solving the lens equation, including the finite lens and source size effects.
By using the evaluated threshold impact parameters, we computed the expected number of microlensing events and gave microlensing constraints on axion clumps with a 95\% confidence level by using the EROS-2 survey and the Subaru HSC observation.
We have also investigated the microlensing constraints on scalar clumps composed of generic light scalar fields with a repulsive self-interaction.

In our analysis, we found that the finite lens size effect can be parametrized by one parameter $w_E$, which is defined by the ratio of the pointlike Einstein ring radius to the typical size of a clump.
In particular, we confirmed that magnifications of source stars are significantly suppressed due to finite extent of a clump for $w_E \lesssim 1$ (see Fig.~\ref{fig:impact parameter}).
As a result, in this parameter region, microlensing events are suppressed, and thus, microlensing constraints become very weak even if clump masses are within targets of the EROS-2 survey and the Subaru HSC observation.
Microlensing constraints on axion clumps from the EROS-2 survey are shown in Figs.~\ref{fig:EROSmF} and \ref{fig:EROS DM fraction}.
In addition to the finite lens size effect, when clump masses are light so that the pointlike Einstein ring radii are comparable to the source star, the finite source size effect becomes very important.
We appropriately included this effect on microlensing constraints when we consider the Subaru HSC observation.
Microlensing constraints on axion clumps from the Subaru HSC observation are shown in Figs.~\ref{fig:SubarumF} and \ref{fig:Subaru DM fraction}.

Assuming the clumps are plentiful in the galaxy, our numerical results showed that the EROS-2 survey and the Subaru HSC observation can constrain clumps whose masses are within $10^{-4}\lesssim M_{\rm clump}/M_{\odot}\lesssim 10$ and $10^{-11}\lesssim M_{\rm clump}/M_{\odot}\lesssim 10^{-5}$, respectively, when axion clumps can be identified with a point lens.
For clumps composed of the ordinary QCD axion, a high breaking scale regime $F_a \gtrsim 10^{12}\,{\rm GeV}$ can be constrained, while a lower breaking scale $F_a \lesssim 10^{12}\, {\rm GeV}$, which corresponds to the traditional axion window cannot be constrained due to the finite source size effect.

Here, we briefly comment about 
 OGLE-IV microlensing survey~\cite{2021ApJS..252...23S,2015AcA....65....1U}.
According to the original analysis~\cite{Niikura:2019kqi}, such a survey provides the tightest microlensing constraint on PBHs in the mass range $[10^{-6}M_{\odot},10^{-3}M_{\odot}]$.
Hence if we include the OGLE-IV survey, we expect that a tighter constraint is obtained for $10^{-6}\lesssim M_{\rm clump}/ M_{\odot}\lesssim10^{-3}$ and $w_E >1$, where the latter condition is necessary to cause microlensing events.
For the case of the QCD axion, such a mass regime requires a large axion decay constant, $F_a \gtrsim 10^{14}\,\text{GeV}$, as shown in Eq.~\eqref{Meq}, which is
ruled out in the standard scenario.
In the present work, we have focused on the EROS-2 and the Subaru HSC surveys. We leave a detailed study of microlensing constraint from the OGLE survey for future study.

A femto-lensing constraint is another appealing experiment, but it was pointed out by the authors of Ref.~\cite{Katz:2018zrn} that the inclusion of the finite source size effect removes this constraint.

\section*{Acknowledgments}
This work was supported by the Academy of Finland Grant No. 318319.
K.F. is supported by JSPS Grant-in-Aid for Research Fellows Grant No. 20J12415.
M. P. H. is supported in part by National Science Foundation Grant No. PHY-2013953. M. P. H. and M. Y. thank the JSPS invitation fellowship. M. Y. is supported in part by JSPS Grant-in-Aid for Scientific Research Grants No. JP18K18764, No. JP21H01080, and No. JP21H00069.\\
$^{1}$\href{mailto:fujikura@resceu.s.u-tokyo.ac.jp}{fujikura@resceu.s.u-tokyo.ac.jp}\\
$^{2}$\href{mailto:mark.hertzberg@tufts.edu}{mark.hertzberg@tufts.edu}\\
$^{3}$\href{mailto:enrico.e.schiappacasse@jyu.fi}{enrico.e.schiappacasse@jyu.fi}\\
$^{4}$\href{mailto:gucci@phys.titech.ac.jp}{gucci@phys.titech.ac.jp}

\appendix

\section{DERIVATION OF THE LENS
EQUATION}\label{app:appendixA}
In this appendix, we derive the lens equation given by \eqref{eq:lens equation} including the extent of the lens.

A microlensing geometrical setup is shown in Fig.~\ref{fig:microlensing setup} of Sec.~\ref{sec:point lens point source}.
From the definition of angles, we obtain
\begin{align}
\beta = \theta - \alpha. \label{eq:lens equation angle}
\end{align}
From the definition of the angular diameter distance, we also obtain 
\begin{align}
D_{\rm LS} \widehat{\alpha} = D_{\rm S}\theta - D_{\rm S}\beta.\label{eq:lens2}
\end{align}
Moreover, the reduced deflection angle $\alpha$ can be expressed by $\widehat{\alpha}$ as 
\begin{align}
\alpha =\frac{D_{\rm LS}}{D_{\rm S}}\widehat{\alpha}. \label{eq:alpha and alphahat}
\end{align}

Let us next estimate the deflection angle $\widehat{\alpha}$.
When a typical radius of an axion clump is much smaller than the length scale of the line of sight such as $D_{\rm S},~D_{\rm L}$ and $D_{\rm LS}$, we can neglect the effect of a lens thickness.
Under this assumption,  one can project the mass distribution onto a plane orthogonal to the line of sight, called the lens plane.
We introduce 3D cylindrical coordinate $(\xi,\chi,z)$, where $\xi,~\chi$, and $z$ represent a radial distance, an azimuthal angle, and a height on the lens plane, respectively.
A surface mass density of the lens projected onto the lens plane, $\Sigma (\xi,\chi)$, is then defined by the following equation:%
\begin{align}
\Sigma  (\xi,\chi) = \int^\infty_{-\infty} \rho(\xi,\chi,z) dz, \label{eq:surface mass density}
\end{align}
where $\rho(\xi,\chi,z) = m_a n(\xi,\chi,z)$ is the energy density of the axion clump.
The total mass of the axion clump on the lens plane within the distance $\xi$ is given by integration of the surface mass density,
\begin{align}
\mathcal{M}(\xi) = 2\pi \int^\xi_0 \Sigma (\xi') \xi' d\xi'. \label{eq:mass projected onto lens plane}
\end{align}
In this calculation, we have assumed that the lens is circular symmetric; {\it i.e.}, the lens object is spherically symmetric.
The deflection angle $\widehat{\alpha}$ is then estimated as $\widehat{\alpha}=4G\mathcal{M}(\xi_L)/\xi_L$, where $\xi_L\equiv D_L \theta$~\cite{Narayan:1996ba}.
This implies that $\mathcal{M}(\xi_L)$ can be regarded as the effective mass within the radius $\xi_L$ causing the gravitational lens.
From Eq.~\eqref{eq:alpha and alphahat}, $\alpha$ can be expressed by
\begin{align}
\alpha(\xi_L) = \frac{D_{\rm LS}}{D_{\rm L}} 4G \frac{\mathcal{M}(\xi_L)}{\xi_L}.
\end{align}
Combining Eqs.~\eqref{eq:lens equation angle} and \eqref{eq:lens2}, we obtain
\begin{align}
\beta (\theta) = \theta - \frac{D_{\rm LS}}{D_{\rm L} D_{\rm S}} \frac{4G\mathcal{M}(\xi_{\rm L})}{\theta}. 
\end{align}
For a pointlike lens, the lens equation is simply given by the replacement with $\mathcal{M(\xi_{\rm L}})\to M$, which is the total mass of the axion clump.

\bibliographystyle{JHEP}
\bibliography{draft}

\end{document}